# The 2012 July 23 Backside Eruption: An Extreme Energetic Particle Event?

N. Gopalswamy

Code 671, Solar Physics Laboratory, NASA Goddard Space Flight Center, Greenbelt, MD 20771

nat.gopalswamy@nasa.gov

S. Yashiro[1], N. Thakur[1], P. Mäkelä[1], H. Xie[1], S. Akiyama[1]

Department of Physics, The Catholic University of America, Washington, DC 20064

Short Title: **An Extreme Energetic Particle Event?**

[1]also at NASA Goddard Space Flight Center, Greenbelt, MD 20771

**ABSTRACT**

The backside coronal mass ejection (CME) of 2012 July 23 had a short Sun to Earth shock transit time (18.5 hours). The associated solar energetic particle (SEP) event had a >10 MeV proton flux peaking at ~5000 pfu, and the energetic storm particle (ESP) event was an order of magnitude larger, making it the most intense event in the space era at these energies. By a detailed analysis of the CME, shock, and SEP characteristics, we find that the July 23 event is consistent with a high-energy SEP event (accelerating particles to GeV energies). The time of maximum and fluence spectra in the range 10-100 MeV were very hard, similar to those of ground level enhancement (GLE) events. We found a hierarchical relationship between the CME initial speeds and the fluence spectral indices: CMEs with low initial speeds had SEP events with the softest spectra, while those with highest initial speeds had SEP events with the hardest spectra. CMEs attaining intermediate speeds result in moderately hard spectra. The July 23 event was in the group of hard-spectrum events. During the July 23 event, the shock speed (>2000 km $s^{-1}$), the initial acceleration (~1.70 km $s^{-2}$), and the shock formation height (~1.5 solar radii) were all typical of GLE events. The associated type II burst had emission components from metric to kilometric wavelengths suggesting a strong shock. These observation confirm that the 2012 July 23 event is likely to be an extreme event in terms of the energetic particles it accelerated.

## 1. Introduction

The backside coronal mass ejection (CME) of 2012 July 23 has received considerable attention because many of its characteristics place it among historical extreme solar events (Baker et al. 2013; Russell et al. 2013; Ngwira et al. 2013; Liu et al. 2014; Liou et al. 2014; Gopalswamy et al. 2014a; Temmer and Nitta, 2015; Joyce et al. 2015; Zhu et al. 2016; Riley et al. 2016, among others). For example, the shock transit time from the Sun to 1 au was ~18.5 hours, similar to the two 2003 Halloween events on October 28 and 29 and to 13 other fast-transit events since the Carrington event in 1859 (Gopalswamy et al. 2005a; Cliver et al. 1990a,b). Based on the measured magnetic field and plasma properties, it was estimated that the CME would have caused a geomagnetic storm comparable to that of the 1859 Carrington storm if it were Earth-directed (Baker et al. 2013; Liu et al. 2014; Gopalswamy et al. 2014a; 2015a). The solar energetic particle (SEP) event associated with the shock was so intense that the shock properties were significantly affected by the nonthermal particles (Russell et al. 2013). Unfortunately, the particle detectors on board the Solar Terrestrial Relations Observatory (STEREO, Kaiser et al. 2008) do not have energy channels above 100 MeV, so we do not know if particles were accelerated to very high energies. The highest-energy particles (>1 GeV) interact with Earth's atmosphere and produce secondaries that reach Earth's surface and hence are called ground level enhancement (GLE) events. GLEs have important space weather consequences because they can significantly contribute to the lifetime radiation dosage to airplane crew and passengers in polar routes (e.g., Shea and Smart 2012). In this sense, the GLE events are considered extreme particle events that occur only about a dozen times during a solar cycle.

SEP events with GeV particles are generally rare. Typically about a dozen events occur during each solar cycle, although only 2 GLEs were reported in cycle 24, probably due to the change in the state of the heliosphere (Gopalswamy et al. 2013a; Thakur et al. 2014; Gopalswamy et al. 2014b). It appears that the 2012 July 23 event would have been another GLE event, if it had occurred on the frontside of the Sun. The purpose of this paper is to examine the event from the perspectives of CME kinematics, SEP intensity and spectrum, and radio-burst association to see if the 2012 July 23 event can be considered as an extreme particle event. The reason for considering these properties is clear from the following facts. Particles up to GeV energies are accelerated by strong shocks driven by CMEs of very high speeds (~2000 km $s^{-1}$) and intense soft X-ray flares (see Gopalswamy et al. 2010; 2012a). The high speed is typically attained very close to the Sun, so the density and magnetic field in the corona is high for efficient particle acceleration (e.g., Mewaldt 2012; Gopalswamy et al. 2014b). The high CME speed implies that a fast mode MHD shock forms close to the Sun as indicated by the onset of metric type II radio bursts, typically at heights <1.5 solar radii (Rs). CMEs attaining high speeds near the Sun have to accelerate impulsively, thus these events are characterized by high initial acceleration (~2 km $s^{-2}$, see Gopalswamy et al. 2012a). This is in contrast to slowly accelerating CMEs (from filament regions outside of active regions) that form shocks at large distances from the Sun and do not accelerate particles to energies more than a few tens of MeV (Gopalswamy et al.

2015b,c). Accordingly, the SEP spectra of such events are very soft, as opposed to the hard spectra of GLE events. Whether an event has a soft or hard spectrum is an important information because the hard-spectrum events have stronger space weather impacts (see e.g., Reames 2013). SEP events with GeV components are accompanied by type II radio bursts from meter (m) wavelengths to kilometer (km) wavelengths (Gopalswamy et al. 2005c; 2010). Type II bursts occurring at such wide raging wavelengths imply strong shocks throughout the inner heliosphere (Gopalswamy et al. 2005b).

## 2. Data Overview

The 2012 July 23 SEP event was detected as a 13-pfu event by GOES particle detectors in the >10 MeV channel (pfu = 1 particle per ($cm^2$ s sr)). The >10 MeV intensity remained at elevated levels for at least 10 days. The >50 MeV and >100 MeV intensities also remained elevated for several days. This is quite a long duration for a single event: the Halloween 2003 events had a similar duration, but had contributions from several eruptions (Gopalswamy et al 2005a; Mewaldt et al. 2005). The 2012 July 23 CME originated on the backside of the Sun from S17W141 from NOAA active region (AR) 11520 that had produced three large SEP events during its disk passage on July 12 (96 pfu), July 17 (136 pfu), and July 19 (70 pfu). CMEs associated with these SEP events were halos (Gopalswamy et al. 2014a; 2015c) and were observed in multiple views from the Solar and Heliospheric Observatory (SOHO) and STEREO. Gopalswamy et al. (2014a) applied the Graduated Cylinder Shell (GCS) Model (Thernisien 2011) to the STEREO and SOHO images of the CMEs, and obtained the peak leading-edge CME speeds as 1415 km $s^{-1}$ (July 12), 1881 km $s^{-1}$ (July 17), and 2048 km $s^{-1}$ (July 19). The CME on July 23 had the highest peak speed: ~2600 km $s^{-1}$. But for the STEREO mission, the July 23 event would have been just an ordinary backside SEP event. The enormous magnetic content of the interplanetary CME (ICME) associated with this eruption (Liu et al. 2014) and the extreme interplanetary (IP) shock properties (Russell et al. 2013) would not have been known.

In addition to the GOES satellites, there were other spacecraft along the Sun-Earth line that observed the SEP event. In particular, we use data from the Solar Anomalous and Magnetospheric Particle Explorer (SAMPEX, Baker et al. 1993). The particle detectors on board STEREO Ahead (STA, W121) and STEREO Behind (STB, E115) also observed the particle event. Clearly STA was magnetically best connected and accordingly detected a prompt increase in particles. Even though poorly connected, spacecraft along the Sun-Earth line and STB detected the particle event, suggesting a longitudinal spread of at least ~245$^{o}$. Using STA's particle detectors we were able to obtain the >10 MeV flux to compare with the GOES data. We also obtain the 10-100 MeV fluence spectrum using SAMPEX and GOES (Sandberg et al. 2014) data to compare with other large SEP events from cycles 23 and 24. We perform both case studies and statistical analyses in assessing the significance of the fluence spectrum of the July 23 event.

Since type II bursts represent electron acceleration by the same shock that also accelerates protons, we consider type II bursts from ground based observatories and from Wind and STEREO. The Radio and Plasma Wave (WAVES) Experiment on board Wind (Bougeret et al. 1995) and STEREO (Bougeret et al. 2008) help us infer shock properties from the dynamic spectra. We also use the Sun Earth Connection Coronal and Heliospheric Investigation (SECCHI, Howard et al. 2008) on board STEREO to study the solar source and CME kinematics. In particular, we use images from SECCHI's Extreme Ultraviolet Imager (EUVI) and the two coronagraphs COR1 and COR2 to track the CME from the solar surface to about 15 Rs. Fortunately, the eruption in question occurred just behind the limb in STB view, so deriving CME kinematics from sky-plane data is accurate. The CME was also well observed in the field of view (FOV) of the Large Angle and Spectrometric Coronagraphs (LASCO, Brueckner et al. 1995) up to ~32 Rs. We combine the white-light observations from the three views to fit a flux rope model to the CME and track it to obtain the height-time history of the CME and its leading shock.

## 3. Analysis and Results

### 3.1 Event Intensity

In order to compare the 2012 July 23 event with the list of large SEP events observed by the GOES spacecraft, we estimate the >10 MeV flux using the STEREO's In-Situ Measurements of Particles and CME Transients (IMPACT, von Rosenvinge et al. 2008) data. In particular, we use data from IMPACT's High Energy Telescope (HET) and Low Energy Telescope (LET) that cover the energy ranges of 13-100 MeV and 3-30 MeV, respectively. We computed the >10 MeV flux in two different ways: (1) we fit a power law to the HET data in the 13-100 MeV range every 5 minutes and extrapolate the spectrum to 10 MeV on the low-energy side and to 150 MeV on the high-energy side. (2) We combine the HET and LET data to plot the spectrum (no fitting) and compute the integral flux from 10 MeV to 150 MeV. In both cases, we assume that the contribution to the >10 MeV flux from particles with energies >150 MeV is negligible.

In order to check whether it is reasonable to use the STEREO spectra for computing the >10 MeV integral flux, we make use of the multi-spacecraft observations of the 2006 December 13 SEP event, which also had a GLE. In particular, STA and STB were very close to Earth on 2006 October 26 because the event occurred at shortly after the STEREO launch. The event was also observed by GOES and SAMPEX, so we can compare the >10 MeV GOES flux with that derived from STEREO data. The SAMPEX data show that there is a spectral break at energies close to 100 MeV. Figure 1 shows the spectra of the 2006 December 13 event near the peak and the >10 MeV flux evolution obtained by obtaining 10-MeV spectrum during every time step. We see that the derived >10 MeV flux agrees quite well with the observed >10 MeV GOES flux within ~4%.

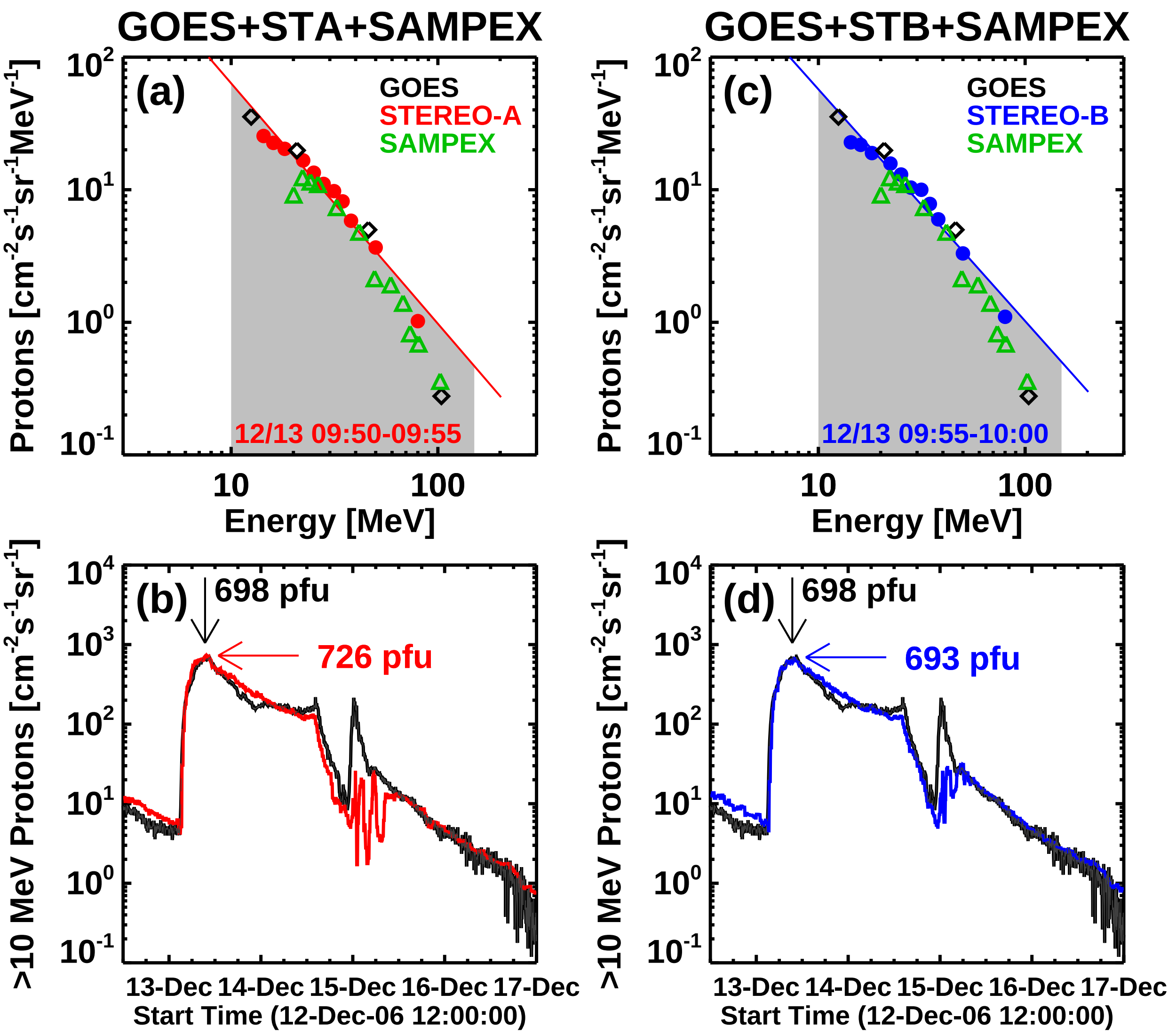

Figure 1. Instantaneous spectrum of the 2006 December 13 SEP event for the interval 19:50 to 19:55 UT (a) constructed from GOES, STA, and SAMPEX data and the corresponding >10 MeV flux (b) derived from a power-law fit to the STA data points in the 10-100 MeV range done every 5 min. In (c) and (d) similar plots are made using STB data. The peak >10 MeV from GOES (black curves), STA (red curves) and STB (blue curves) are shown. Note that the peak >10 MeV fluxes derived from STA and STB agree quite closely with the GOES >10 MeV flux.

Figure 2 shows the >10 MeV integral proton fluxes (5-minute averages), along with examples of the STEREO spectra of the July 23 event. There was an intense energetic storm particle (ESP) event when the CME-driven shock arrived at STA. The peak ESP flux was 64055 pfu, which is more than an order of magnitude larger than the peak SEP flux (5528 pfu) as can be seen in Fig. 2b. The peak SEP flux was attained at 13:57:30 UT (in the interval 13:55-14:00 UT). We also combined the HET and LET data to plot the spectrum (no fitting) to compute the >10 MeV integral flux (see Fig. 2c). In this case, we used the last two data points to linearly extrapolate the flux values to 150 MeV. We see from Fig. 2d that the peak SEP flux (4643 pfu) and ESP flux (43766 pfu) are smaller by 16% and 32% than those obtained from HET data alone.

Since the July 23 event had a large ESP increase, we wanted to compare it with other events with similar increase. We also wanted to compare with GLE events with the highest percentage increase in intensity because we would like to know if high-energy particles were accelerated during the July 23 event. ESP events are typically associated with eruptions close to the disk center and show significant east-west hemispheric asymmetry (e.g., Sarris et al. 1984; Reames 1999; Mäkelä et al. 2011). In the eastern hemisphere, ESPs occur in events originating all the way to the east limb. When we examined the source regions of ESP events listed in the CDAW catalog (http://cdaw.gsfc.nasa.gov/CME_list/sepe/), we found that the events were confined to <W35 in cycles 23 and 24, while they occurred all the way to east limb. The average longitudes were E28 and W18. Since the July 23 CME originated close to the central meridian in STA view, that spacecraft observed the ESP event. We consider four ESP events from the past that had >10 MeV ESP flux exceeding 30,000 pfu. We also consider the top four GLE events in terms of percentage increase above the background (>90%) as observed by the Oulu Neutron Monitor (http://cosmicrays.oulu.fi/GLE.html, Usoskin et al. 2001).

Table 1: Intense ESP and GLE events since 1976

| Date | SEP Flux (pfu) | ESP Flux (pfu) | GLE (%)[b] | Source | Ref |
|---|---|---|---|---|---|
| 1989 Oct 19 | 3000 | 40,000 | 37 | S27E10 | 1 |
| 1991 Mar 23 | 4570 | 43,000 | N | S26E28 | 2 |
| 2001 Nov 4 | 4340 | 31,700 | 3 | N06W18 | 3,4 |
| 2003 Oct 28 | 4940 | 33,600 | 5 | S16E08 | 3,4 |
| 1989 Sep 29 | 3100 | ---- | 174 | S25W103 | 5 |
| 1989 Oct 24 | 2200 | ---- | 94 | S30W55 | 1 |
| 2005 Jan 20 | 1860 | ---- | 269 | N12W58 | 3,4 |
| 2006 Dec 13 | 700 | ---- | 92 | S06W23 | 3,4,6 |
| 2012 Jul 23 | 5528 | 64,055[a] | ? | S17W20[c] | 7 |

**Notes.** [a]based on spectra obtained from HET data; the SEP and ESP peak values are 4643 and 43,766 pfu, respectively when spectra were obtained from LET and HET data; the SEP and ESP are 1-2% different from what was published in Gopalswamy et al. (2014b) that did not exclude the 14.35 MeV data point. Mewaldt et. al. (2013) obtained an ESP peak of 35,800 pfu by combining LET and HET data, fitting a Band function to the combined data, and integrating the fitted spectrum over the range 10-100 MeV. Their value is smaller than ours (43,766 pfu) by 18%. The difference reduces to 9% when we restrict to 10-100 MeV range. The remaining difference is most likely due to the spectral fit they used (we did not fit), different way of treating HET-LET differences, and the fact that they took the peak value to occur five minutes before ours. [b]The percentage ground level enhancement from the Oulu Neutron Monitor (http://cosmicrays.oulu.fi/GLE.html); N indicates no GLE observed. "?" in the 2012 July 23

event is to indicate that we are trying to find if GeV particles were accelerated during this event. [c]The source location in STA view; it is S17W141 in Earth view. **References.** (1) Reeves et al. (1992); Lario et al. (2001); (2) Shea and Smart (1993); Kocharov et al. (1995); (3) Gopalswamy et al. 2010; 2012a; (4) Mewaldt et al. (2012); (5) Cliver et al. (1993); Miroschnichenko et al. (2000); (6) Liu et al. (2008); (7) Gopalswamy et al. (2014a)

Table 1 lists these ESP and GLE events along with the >10 MeV SEP fluxes, which were in the range 2000 to 5000 pfu. The SEP fluxes were measures as the peak value before the onset of the associated ESP event. Even though there is a weak correlation between the SEP flux and the GLE intensity (Gopalswamy et al. 2012a; Thakur et al. 2016), there are a couple of significant exceptions: the 1991 March 23 event was the largest ESP event with high SEP flux in Table 1, but it was not associated with a GLE (Shea and Smart 1993), and the second largest GLE event in solar cycle 23, viz., the 2006 December 13 had an SEP flux of only ~700 pfu (Liu et al. 2008). The 1991 March 23 event was associated with an X9.4 flare from AR 6555 that started the previous day around 22:43 UT. Although there was no CME observations for this event, there was a Moreton wave with a speed of ~1900 km $s^{-1}$ (Kocharov et al. 1995) and an intense metric type II radio burst reported in the Solar Geophysical Data. These observations indicate the presence of a high-speed CME. The CME-driven shock resulted in the largest sudden commencement since 1986 (Araki 2014). It appears that the CME had slowed down considerably because the estimated transit time from the flare onset (22:43 UT) to the sudden commencement (3 UT on March 24) was ~28 hours (transit speed ~1500 km $s^{-1}$). There was also an intense Forbush decrease consistent with the shock arrival at 1 au (Shea and Smart 1993). The flare was also associated with neutron and gamma-ray continuum emissions (Kocharov et al. 1995).

Table 1 shows that among the four ESP events, three were associated with GLEs and one was not (1991 March 23); the March 23 event was the easternmost of the four. We will use this event for a detailed comparison with the July 23 event. Among the four GLE events in Table 1, three had better connectivity than the July 23 event and had similar connectivity as the 2006 December 13 event. The 2003 October 28 GLE event had a shock transit time of ~18.9 hours, very similar to that of the 2012 July 23 event. None of the other events had such short transit times.

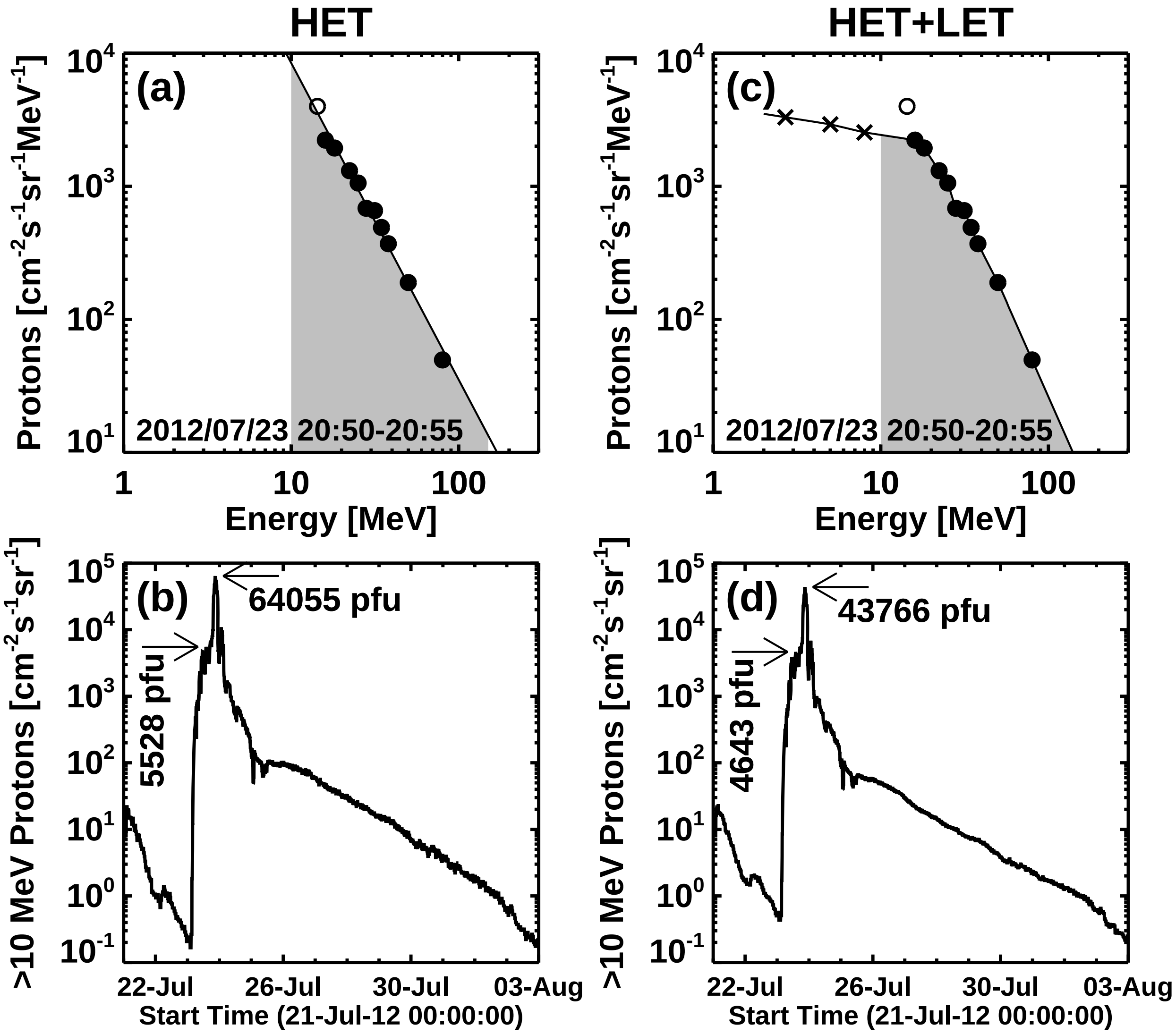

Figure 2. (a) Instantaneous proton spectrum obtained using STEREO/IMPACT HET data during 20:50 to 20:55 UT on 2012 July 23 and (b) the corresponding >10 MeV proton flux as a function of time. (c) Proton spectrum when both HET and LET data are combined and (d) the corresponding >10 MeV intensity. The crosses represent LET data points in the three channels: 1.8 - 3.6 MeV, 4.0 - 6.0 MeV, and 6.0 - 10.0 MeV. The data in the highest LET channel are not available. In computing the >10 MeV intensities, the shaded area under the curves are summed over time (using 5-min data). The open circle represents the data point at 14.35 MeV, which seemed to be an overestimate for a period of ~1 h near the maximum. Therefore, we excluded this data point in computing the integral flux in (c) and (d). In (a) the spectrum is extrapolated to 10 MeV on the low energy side and up to 150 MeV on the high energy side. In (c), the data point corresponding to the lowest HET channel is connected to that in the highest LET channel to identify where the connecting line crosses 10 MeV.

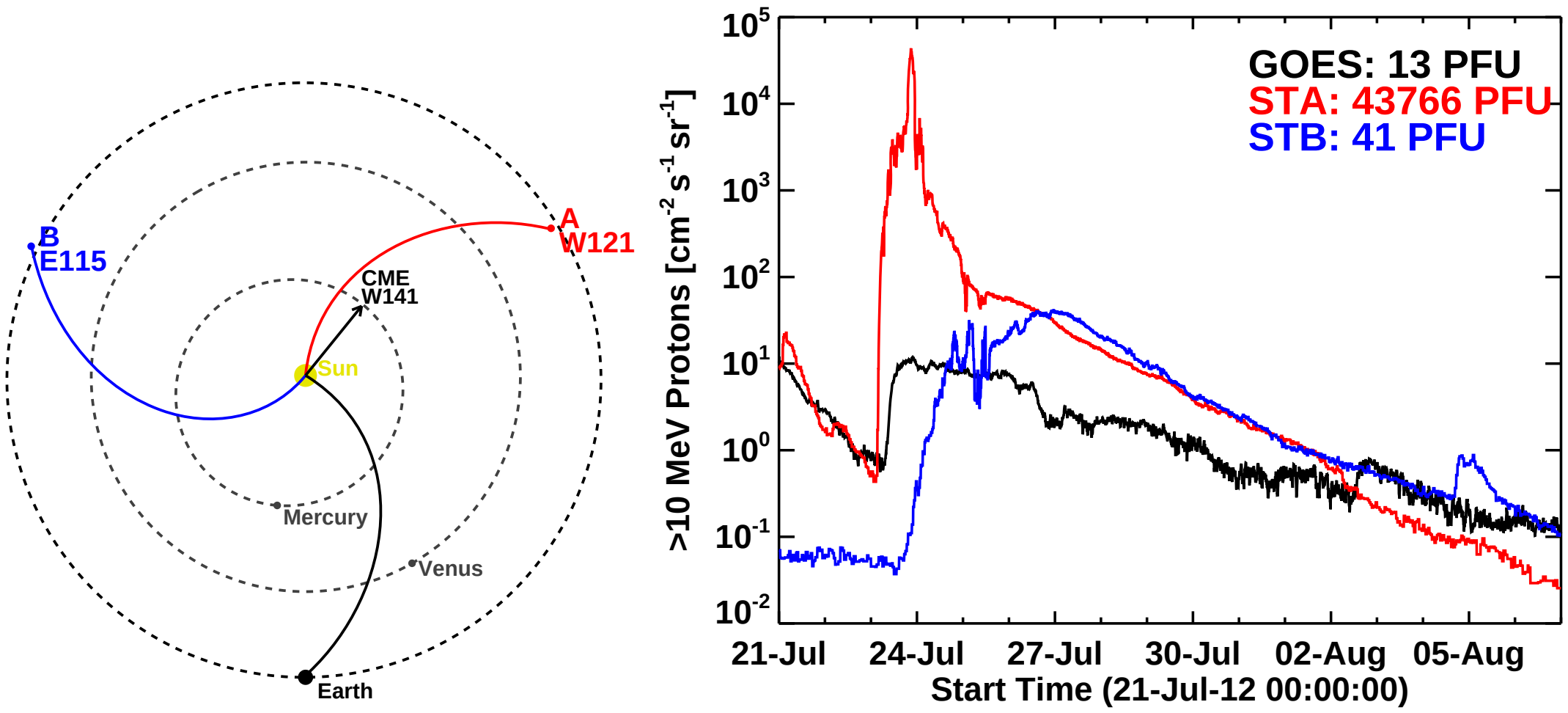

Figure 3. (left) location of STA (W121), STB (E115), with respect to Earth. The direction of the CME (W141) is shown. The locations of Mercury and Venus are also shown. The dashed circles represent the orbits of the three planets. The Parker spirals are based on the solar wind speeds measured at STA, Earth, and STB. (right) Extended time – proton intensity plot of the SEP event. The GOES-equivalent >10 MeV flux detected by STA and STB was computed by obtaining proton spectra from STEREO LET and HET during every time step in the 10-100 MeV range and assuming that no contribution from particles at energies >150 MeV. The peak values of >10 MeV proton flux are noted on the plot as detected at GOES, STA, and STB.

Figure 3 compares the >10 MeV SEP flux at three locations: STA, STB, and Earth (GOES). The >10 MeV flux at STA and STB flux were computed using the LET+HET spectra. The poorly-connected locations (STB, GOES) have two orders of magnitude lower flux values. The event onset was also delayed by several hours at Earth and STB. The SEP event is likely to be circumsolar because particle data from the Mercury Surface, Space Environment, Geochemistry, and Ranging mission (MESSENGER, Zurbuchen et al. 2011) show clear enhancement in SEP intensity in the energy channels >841 keV. MESSENGER was located to the east of the Sun-Earth line. Figure 4 shows the ion count rate as a function of time in three energy channels (0.841 to 2.065 MeV, 2.065 to 2.750 MeV, and the unbounded). It appears that the onset of the event is around 9 UT, between the onsets at Earth and STB.

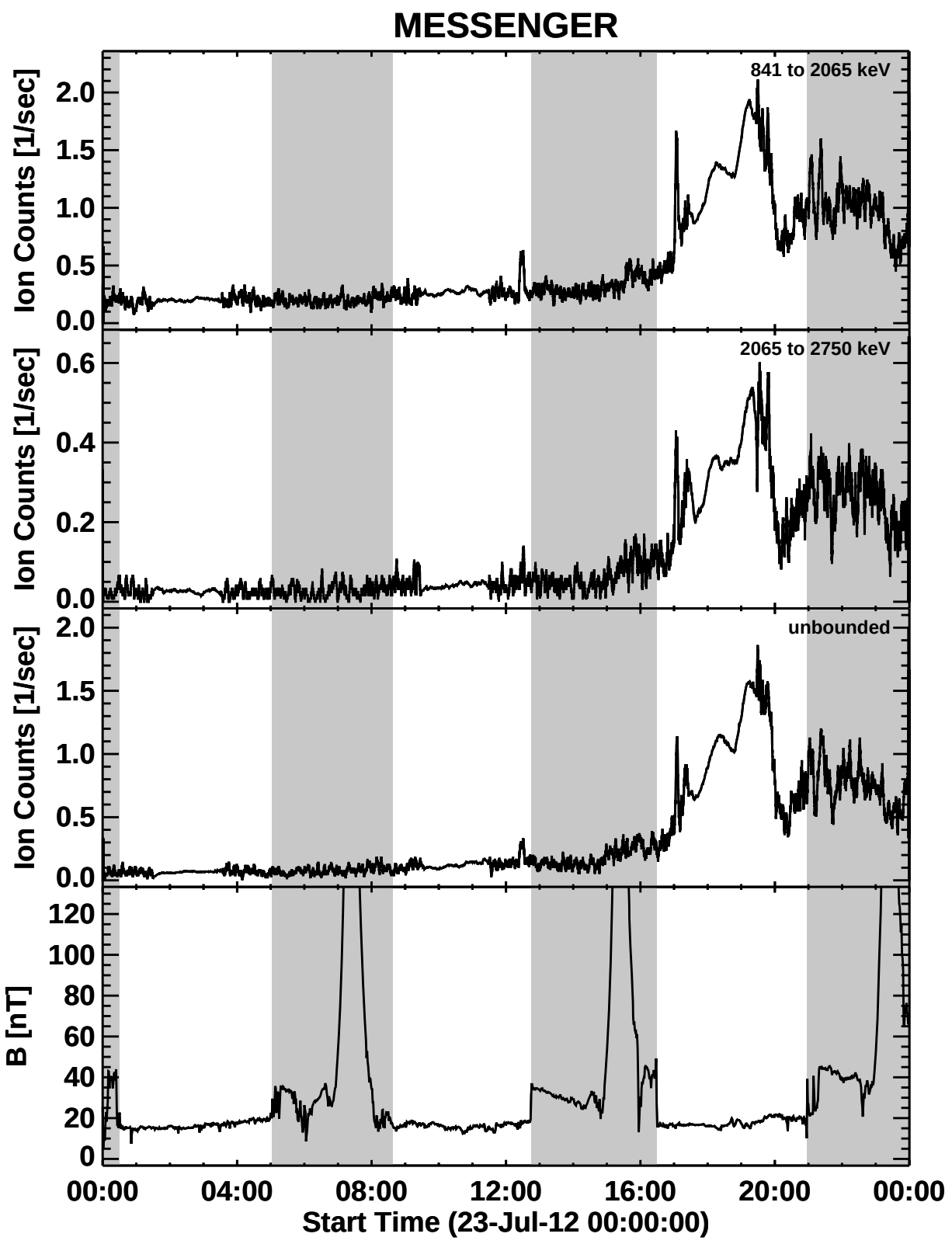

**Figure 4.** Top three panels: SEPs measured by the MESSNGER particle detectors in three energy channels. Bottom panel: The local magnetic field strength B in nT. The shaded area corresponds to the duration the MESSENGER spacecraft was inside Mercury's magnetosphere.

### 3.2 Time of Maximum (TOM) Spectra

A recent investigation of SEP events associated with filament eruptions (FEs) outside active regions showed that the Time of Maximum (TOM) spectra are extremely steep suggesting that particles are not accelerated to energies greater than several tens of MeV (Kahler 2001; Gopalswamy et al. 2015b). The 1-100 MeV TOM spectra for these events had a spectral index $\gamma \geq 4$. The spectra were obtained from the proton intensity data from Energetic and Relativistic Nuclei and Electron (ERNE; Torsti et al. 1995) instrument on board SOHO. The spectral shape was assumed to be a power law of the form, $dN/dE = kE^{-\gamma}$ where N and E are the energetic particle density and energy, respectively, k is a constant, and $\gamma$ is the spectral index. The intensity maximum was taken at a time well before the ESP event (if present). The power-law spectral index $\gamma$ ranged from 4.15 to 4.69. These values corresponded to the upper end of the approximate range of $2 < \gamma < 4.5$ for well-connected >10 pfu SEP events (Kahler 2001). One of the important outcomes of this study was that there were events with $\gamma > 4$ even originating from active regions. What was common to all the $\gamma > 4$ events was the large height of shock formation as indicated by the low starting frequency of the associated type II radio bursts. The type II radio bursts typically started in the Wind/WAVES spectral domain (<14 MHz), indicating a shock formation height >2.0 Rs, in contrast to the GLE events in which the shock formation is about 1.5 Rs. For events with harder spectra (γ ~2) particles are accelerated to higher energies.

Figure 5 shows that the TOM spectrum of the July 23 event obtained from STA/HET data is very hard, with γ~1.30. This is a good indication that particles were accelerated to high energies. We have also shown the TOM spectrum of the 1991 March 23 event, which had comparable SEP and ESP proton flux in the >10 MeV channel. The 10-100 MeV spectral index for this event is γ ~1.80, higher than that of the July 23 event. The spectral indices are much smaller than those of the FE SEP events (γ>4), suggesting that the TOM spectrum of the 2012 July 23 event is much harder. The rollover at higher energies is pronounced in the March 23 event. Recall that the March 23 was an eastern event and poorly connected to the source. The TOM spectral indices of the July 23 event from GOES (1.74) and STB (3.0) are also higher than that from STA (1.30) most likely because GOES and STB are poorly connected to the source region.

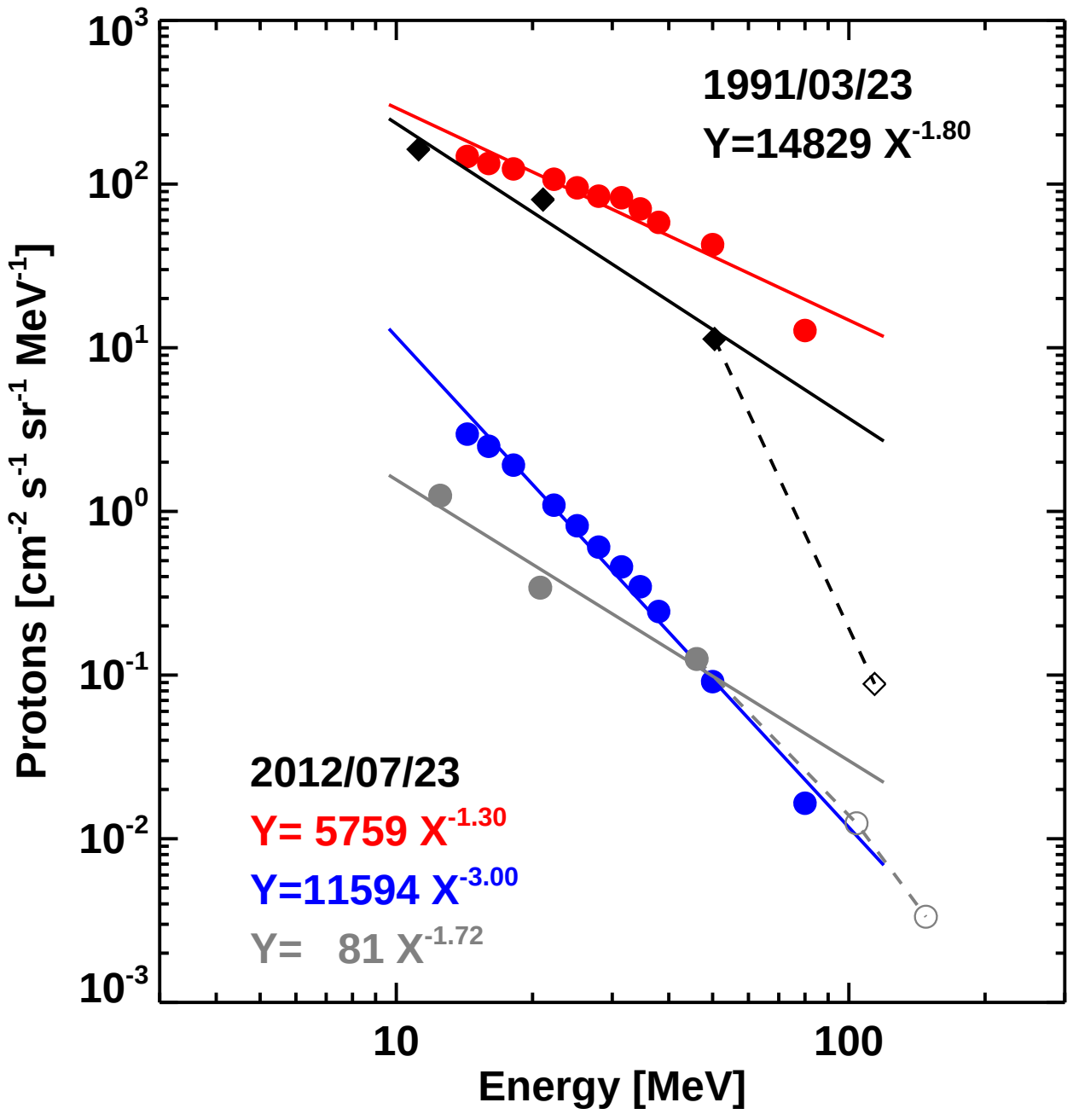

Figure 5. TOM spectrum of the 2012 July 23 event from STA data (red), STB (blue), and GOES (gray) compared with that of the 1991 March 23 event (black), which is one of the largest SEP events since 1976 detected by GOES (see Table 1). Clearly there is a roll over at higher energies in both events, but more pronounced in the 1991 event. Power-law fits in the 10-100 MeV range are shown on the plot, with indices of 1.30, 3.00, and 1.74 from STA, STB, and GOES data, respectively for the 2012 July 23 event and 1.80 for the 1991 March 23 event from GOEs data. The dashed lines connecting to the data point beyond 100 MeV (open symbols) indicate significant rollover.

For events with very high particle fluxes, the ERNE instrument saturates, so it is difficult to use ERNE data in intense SEP events. Fortunately, Gopalswamy et al. (2015b) found that the ERNE intensities are generally consistent with GOES for the FE SEP events. Therefore, we use GOES data for the events considered in this paper. For the 2012 July 23 event, we use GOES (Earth), STA, and STB data to derive the TOM spectrum.

### 3.3 Fluence Spectra

TOM spectra are somewhat difficult to obtain when there are large fluctuations in the time profiles in various energy channels, making it difficult to identify the peak intensities. Therefore, we consider fluence spectra from GOES and SAMPEX. We chose the starting and ending times of the interval over which the fluence was computed as follows: the starting time was taken as the onset time in the highest channel, while the ending time is taken as the time when the SEP intensity in the lowest channel decayed to the background level. As before, the spectra were obtained in the 10-100 MeV range. SAMPEX data have more channels, so the spectra are more reliable. However, the absolute flux values are uncertain, so we normalize them to the GOES values using channels of overlapping energies (Mewaldt et al. 2012). SAMPEX data are available for the whole of cycle 23, so we use these data to obtain the fluence spectra. In cycle 24, we use GOES data. We group the SEP events into FE SEP events, regular SEP events, and GLE events in comparing their spectra with that of the July 23 event. We restrict the regular SEP events to those originating from the western hemisphere of the Sun. Hereafter, "spectral index" refers to the index of the fluence spectrum.

#### 3.3.1 Case Studies

Figure 6a shows the 10-100 MeV fluence spectrum of the 2012 July 23 event from STA, STB and GOES in comparison with the spectrum of the 1991 March 23 event from GOES. The spectral indices are comparable: 2.28 (STA, 2012 July 23) and 2.24 (GOES, 1991 March 23). The spectral index of the July 23 event from STB (3.15) data is large because the source region was poorly connected to STB. On the other hand, the spectral index is small (1.74) from GOES (also poorly connected to the source), but the data point beyond 100 MeV shows a steep decline. The GOES spectra of the March 23 and July 23 events suggest a rollover beyond 100 MeV most likely due to poor longitudinal connectivity. Before a statistical analysis, we compare the fluence spectrum of the 2012 July 23 SEP event with that of a few SEP events from cycles 23 and 24 (Fig. 6b,c): the two GLE events of cycle 24 (2012 May 17 and 2014 January 6), the 2006 December 13 GLE (also observed by STEREO), the 2005 January 20 GLE (the largest GLE of cycle 23), the two SEP events associated with filament eruptions outside active regions (on 2011 November 26 in cycle 24 and 2004 April 11 in cycle 23), and finally two regular well-connected large SEP events (2001 April 14 in cycle 23 and 2012 January 27 in cycle 24). For comparisons with other events, we use the STA spectral index for the July 23 event because STA is better connected to the source region.

The spectral index of the 2012 July 23 event is close to that of the four GLE events in Fig. 6. Note that the 2006 December 13 event was observed by particle detectors on multiple spacecraft: STA, STB, SAMPEX, and GOES. The spectral indices obtained from STB (γ=2.05) and SAMPEX (γ=2.07) are consistent with the STA index (γ=2.02). SAMPEX also had data in higher energy channels, which when included result in slightly steeper slopes (γ=2.48) because of the roll over beyond ~100 MeV (the last three SAMPEX data points clearly show the roll over). In the case of the 2005 January 20 GLE, the 10-100 MeV spectral index (2.13) is very similar to the one using the full energy range (2.17) because the roll over is not significant. The indices of the cycle-24 GLE spectra are nearly the same: 2.48 and 2.54. The fluence spectrum of the 1989 October 19 event (also a GLE – see Table 1) was found to be identical to that of the July 23 event (Mewaldt et al. 2016, private communication). The spectral indices of the GLE events and the July 23 event are clearly much smaller than that of the two FE SEP events (4.50 and 5.46). The two regular SEP events had similar 10-100 MeV spectral indices that are intermediate between the GLE and FE SEP events.

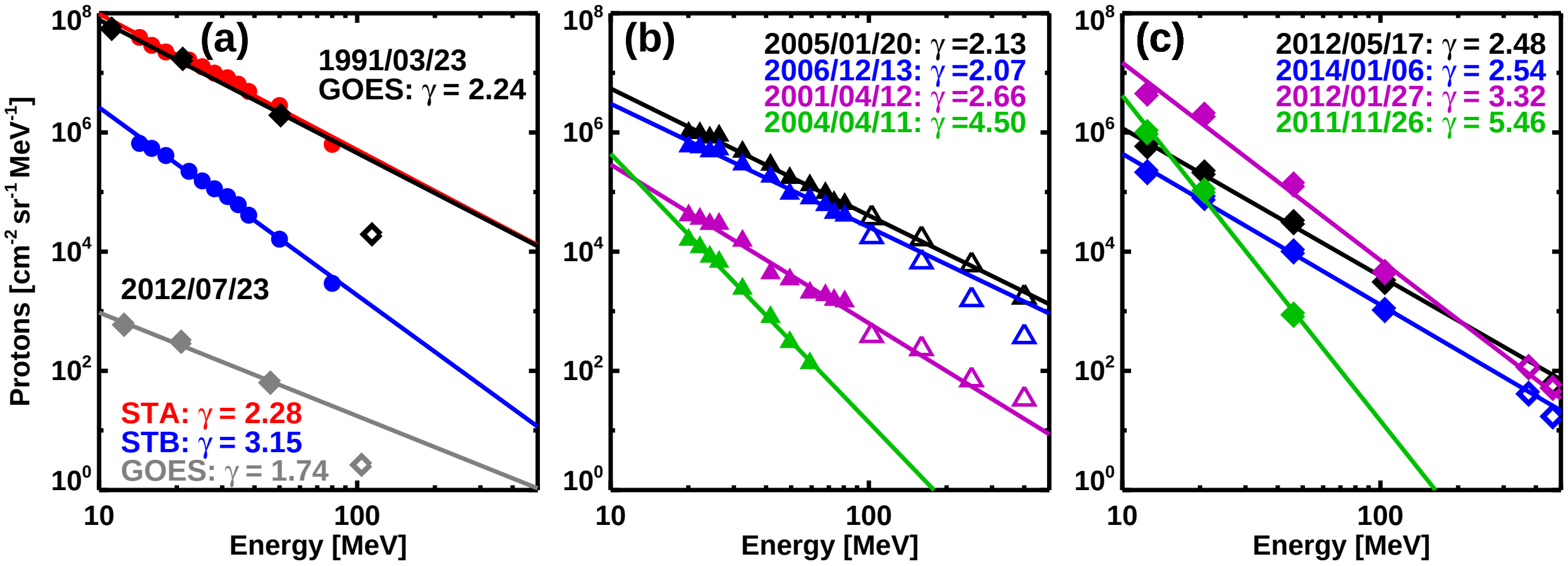

Figure 6. STA (HET) fluence spectrum of the 2012 July 23 event compared with that of several large SEP and GLE events. (a) **Spectra of the 2012 July 23 event from STA, STB, and GOES compared with that of the 1991 March 23 event.** (b) Fluence spectra of a set of cycle-23 SEP events including two GLEs (2006 December 13 and 2005 January 20), a regular SEP event (2001 April 12), and a FE SEP event (2004 April 11). (c) Fluence spectra of a set of cycle-24 SEP events including two GLEs (2012 May 17 and 2014 January 06), a regular SEP event (2012 January 27), and a FE SEP event (2011 November 26). The spectral index was calculated using the data points represented by solid symbols; the open symbols indicate the data availability and possible rollovers. Significant rollovers can be seen in the case of the 1991 March 23 and 2006 December 13 events.

### 3.3.2 Statistical Comparisons

In order to confirm the results shown in Fig. 6, we computed the fluence spectra of all GLE events, FE SEP events and well-connected regular SEP events in cycles 23 and 24. We dropped three regular western SEP events (2012 May 26, 2012 June 14, and 2015 June 25) because there were only two data points to derive the fluence spectrum. Table 2 lists the dates and times of the SEP onset in columns 1 and 2. Columns 3-6 list the SEP type, fluence spectral index γ, **the error in γ,** and the energy range over which γ was computed. The CME first appearance time in the LASCO FOV is given in column 7. Column 8 lists the CME initial speed obtained from the first two height-time measurements obtained in the LASCO/C2 FOV. The sky-plane speed from the SOHO/LASCO CME catalog and the space speed (deprojected speed) are listed in columns 9 and 10, respectively. The projection correction was made using the cone model for eruptions origination within 60º in longitude from the disk center. For eruptions within 30º from the limb, we used a simple geometrical deprojection. The source location (heliographic coordinates) is given in column 11, followed by flare class (column 12), start time (column 13) and peak time (column 14). Finally, we list the average initial mass acceleration obtained by dividing the space speed by the flare rise time (start to peak in soft X-rays), as described in Zhang & Dere (2006). As in Fig. 6, the cycle-23 spectra are from SAMPEX, while those from cycle 24 are from GOES.

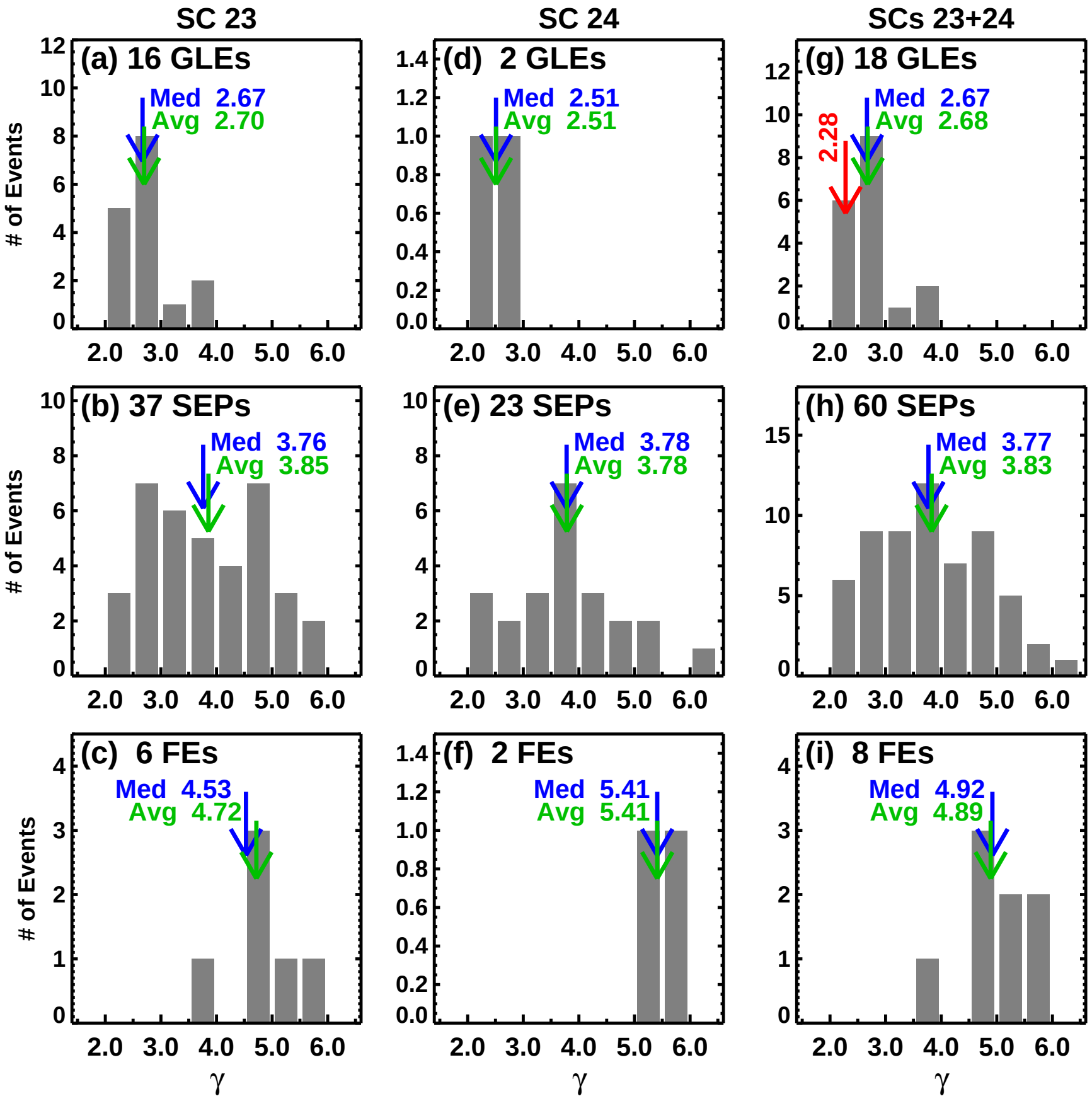

Figure 7. Distributions of 10-100 MeV spectral indices of SEP events in cycle 23 (a-c), cycle 24 (d-f), and the combined set (g-i). GLE events, regular SEP events, and FE SEP events are shown separately. The average (Avg) and median (Med) values of the distribution are marked on the plots. The spectral index of the 2012 July 23 event (γ=2.28) is shown on the GLE plot in the last column for comparison.

Figure 7 shows the distributions of γ for the three groups of SEP events in cycles 23, 24, and the combined set. GLEs have the smallest spectral indices (hardest spectra – average γ=2.68). FE SEP events have the largest indices (softest spectra, average γ=4.89). The well-connected regular SEP events have intermediate values (average γ=3.83). Clearly, the 2012 July 23 SEP event has its γ=2.28 similar to the GLE group, suggesting GeV particles were likely accelerated in this event. The histograms also show a systematic difference in γ as one goes from the GLE events to the regular SEP events and to the FE SEP events. There are many regular SEP events with γ > 4. Some of these are likely to be due to CMEs behaving similar to those in FE SEP events in that they typically have positive acceleration in the coronagraph FOV resulting in shock formation at larger distances. Several such events were also reported in Gopalswamy et al. (2015a). It is also possible that some are indeed FE SEP events, but were not identified as such.

### 3.3.3 Implications: Hierarchical Relationship

In SEP events with GeV particles, the shock forms very close to the Sun, about half a solar radius above the solar surface. In events associated with filament eruptions outside active regions, the shock forms at much larger heights – either in the outer corona or in the interplanetary medium (Mäkelä et al. 2015). Particles can be accelerated efficiently to high energies when shocks form close to the Sun because of the high ambient magnetic field, thus yielding a hard spectrum. The low shock formation height also implies that the associated CMEs accelerate impulsively (initial acceleration is ~2 km $s^{-2}$) to reach supermagnetosonic speeds. On the other hand, in FE SEP events, particles are not accelerated to high energies, consistent with shock formation at large distances from the Sun and the soft spectrum (Gopalswamy et al. 2015a). The regular SEP events show intermediate behavior in shock formation height, initial acceleration, and spectral hardness. Thus, there is clear hierarchical relationship between CME kinematics and the spectral hardness of SEP events.

One way to see the effect of early acceleration is to compare the speeds of CMEs near the Sun. For this purpose, we consider speeds attained by CMEs associated with the SEP events ($V_{in}$ listed in Table 2) estimated from the first two height-time data points listed in the SOHO/LASCO CME catalog (http://cdaw.gsfc.nasa.gov, Yashiro et al. 2004; Gopalswamy et al. 2009a) in the FOV of the C2 telescope (2.5 to 6 Rs). $V_{in}$ was then deprojected by multiplying it by the ratio $V_{Sp}/V_{Sky}$, where $V_{Sp}$ and $V_{Sky}$ are the space speed and the sky-plane speed, respectively. Figure 8 shows a number of scatter plots among γ, the initial CME speed, and the source longitude. It is clear that the GLE and FE-SEP populations are distinct, but the regular SEP population has a wide spread. The FE SEP events occupy the upper left part of the plots in the upper panel, while the GLE events occupy the lower right part; there is no overlap between the two populations. One can also see a trend of lower initial CME speeds in the FE SEP events than in the GLE events. There is a large scatter because other source and environmental factors come into play in addition to the projection effects (see e.g. Gopalswamy et al. 2014a). Despite the large scatter, the correlations are significant because the critical values of the Pearson Correlation coefficients are well below the correlation coefficients shown in Fig. 8(a-c). Another way to see the trend is to average the speeds in each population as shown in Table 3. There is a clear anti-correlation between <V> and <γ>: As one goes from FE SEP events to GLE events, the CMEs get faster and the fluence spectrum becomes harder. The three sets of data pints in the last two columns of Table 3 fit to a regression line, $<\gamma> = - 0.0021<V> + 7.42$ with a correlation coefficient of 0.995.

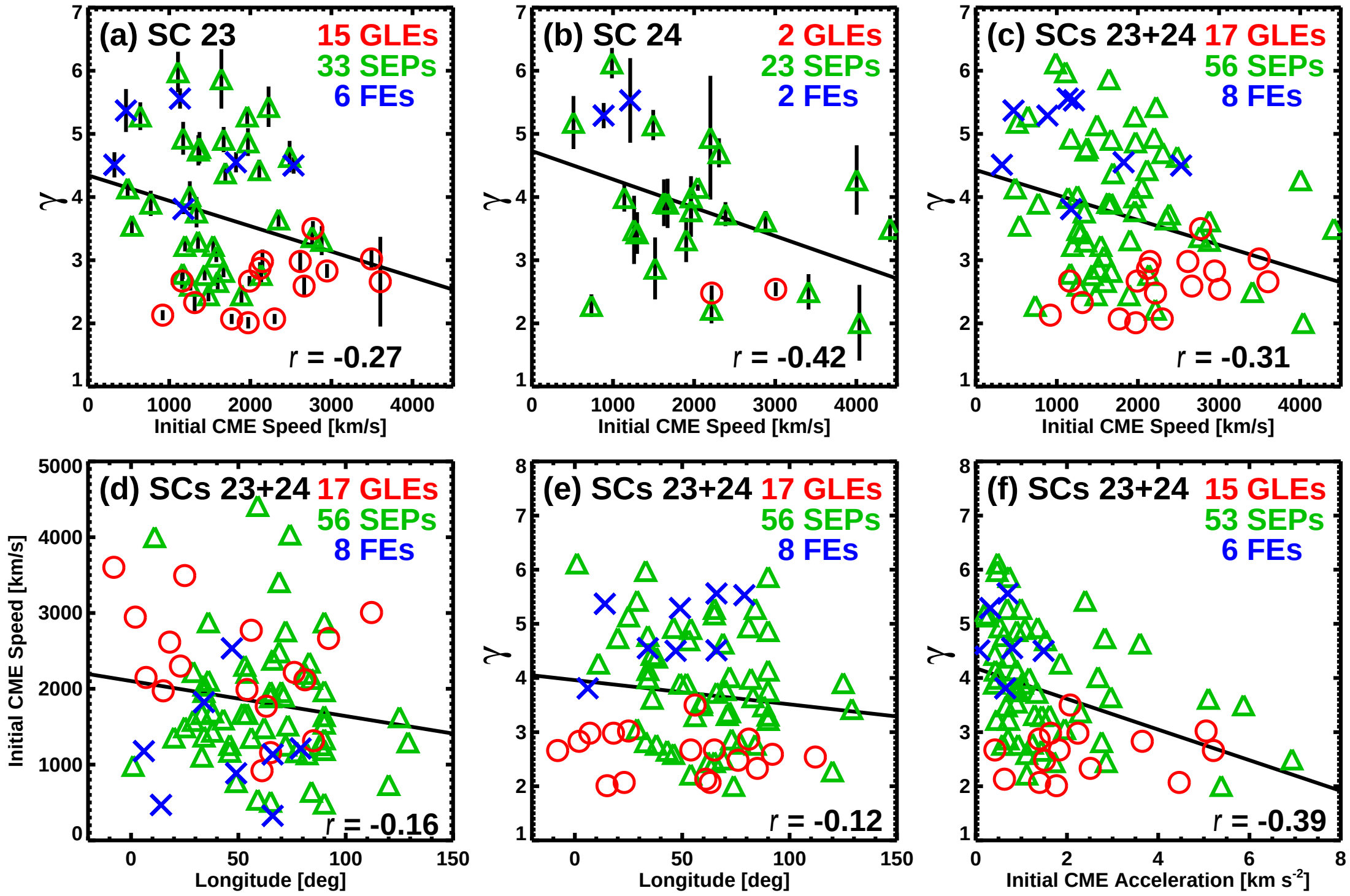

Figure 8. Upper panel: Scatter plots of γ versus the initial CME speed: (a) cycle 23, (b) cycle 24, and (c) the combined set. FE SEP events (blue crosses), regular SEP events (green triangles), and GLE events (red circles) are distinguished. The error bars are from Table 2. The initial CME speed was computed from the first two height-time data points in the LASCO/C2 field of view and deprojected (see text). There are only 15 GLE events plotted in SC 23 because there were no white-light observations for the 1998 August 24 GLE due to a LASCO data gap. The regression line and the correlation coefficients are shown on the plots. The critical values of the Pearson correlation coefficients ($r_c$) are 0.227, 0.301, and 0.180 for the data in (a), (b), and (c), respectively at 95% confidence level. Lower panel: (a) Scatter plot between the CME initial speed and source longitude showing no correlation ($r = -0.16$; $r_c = 0.185$), (b) Scatter plot between the fluence spectral index and source longitude showing no correlation ($r = -0.12$; $r_c = 0.185$) , (c) Scatter plot between the fluence spectral index and initial CME speed showing significant correlation ($r = - 0.39$; $r_c = 0.192$).

Table 3. Average speeds (<V>) and average spectral indices (<γ>) for the three SEP groups

| SEP group | <V> (23) | <γ> (23) | <V> (24) | <γ> (24) | <V> (23+24) | <γ> (23+24) |
|---|---|---|---|---|---|---|
| FE | 1243 km $s^{-1}$ | 4.72 | 1049 km $s^{-1}$ | 5.41 | 1195 km $s^{-1}$ | 4.89 |
| Regular SEP | 1579 km $s^{-1}$ | 3.92 | 2086 km $s^{-1}$ | 3.78 | 1787 km $s^{-1}$ | 3.86 |
| GLE | 2253 km $s^{-1}$ | 2.63 | 2612 km $s^{-1}$ | 2.51 | 2295 km $s^{-1}$ | 2.61 |

Figures 8(d-e) show the dependence of the initial CME speed and γ on the source longitude. Since the initial speed is corrected for projection effects, there is no correlation. The correlation coefficient r = -0.16 is well below $r_c$ = 0.185, suggesting that there is no significant correlation. Similarly, there is no correlation between γ and source longitude (r = -0.12; $r_c$ = 0.185). When we plotted γ against the distance from the well-connected longitude (W45 to W55), we did not find any correlation. The only longitudinal effect we see is in Fig. 8e: in the lower left corner of the plot (longitude <W30 and γ <3.2) we only see GLE events, suggesting possible softer spectra when poorly connected. Recall that we have used only western SEPs, so the effect of the longitudinal connectivity is not clearly seen. Kahler (2001) found a weak correlation (r =0.21) between TOM spectral index and source longitude, but he used eastern events also. While the slope of the regression line in Fig. 8e has a similar trend, the correlation is not significant. When eastern events are included, one expects a noticeable difference as in the individual cases shown in Fig. 6a. When we measured the CME widths at solar particle release in GLE events occurring near the limb, we found that all he widths are >80º. Large longitudinal widths can blur any longitudinal dependence of γ. The widths can also be different depending on the orientation of the CME (Gopalswamy et al. 2015c). Results of such an analysis involving eastern events, CME widths, and orientations will be reported elsewhere.

In Fig. 8f, we show that the correlation between γ and initial CME acceleration derived from CME space speed and flare rise time is significant (r = -0.39; $r_c$ = 0.192 at 95% confidence level), confirming the result in Fig. 8c. The correlation coefficients are similar, justifying the use of the initial CME speed as a proxy to the initial acceleration. The main difference is that the initial CME speed can be obtained from coronagraph observations alone, while the initial acceleration needs to be computed using coronagraph observations and flare rise time.

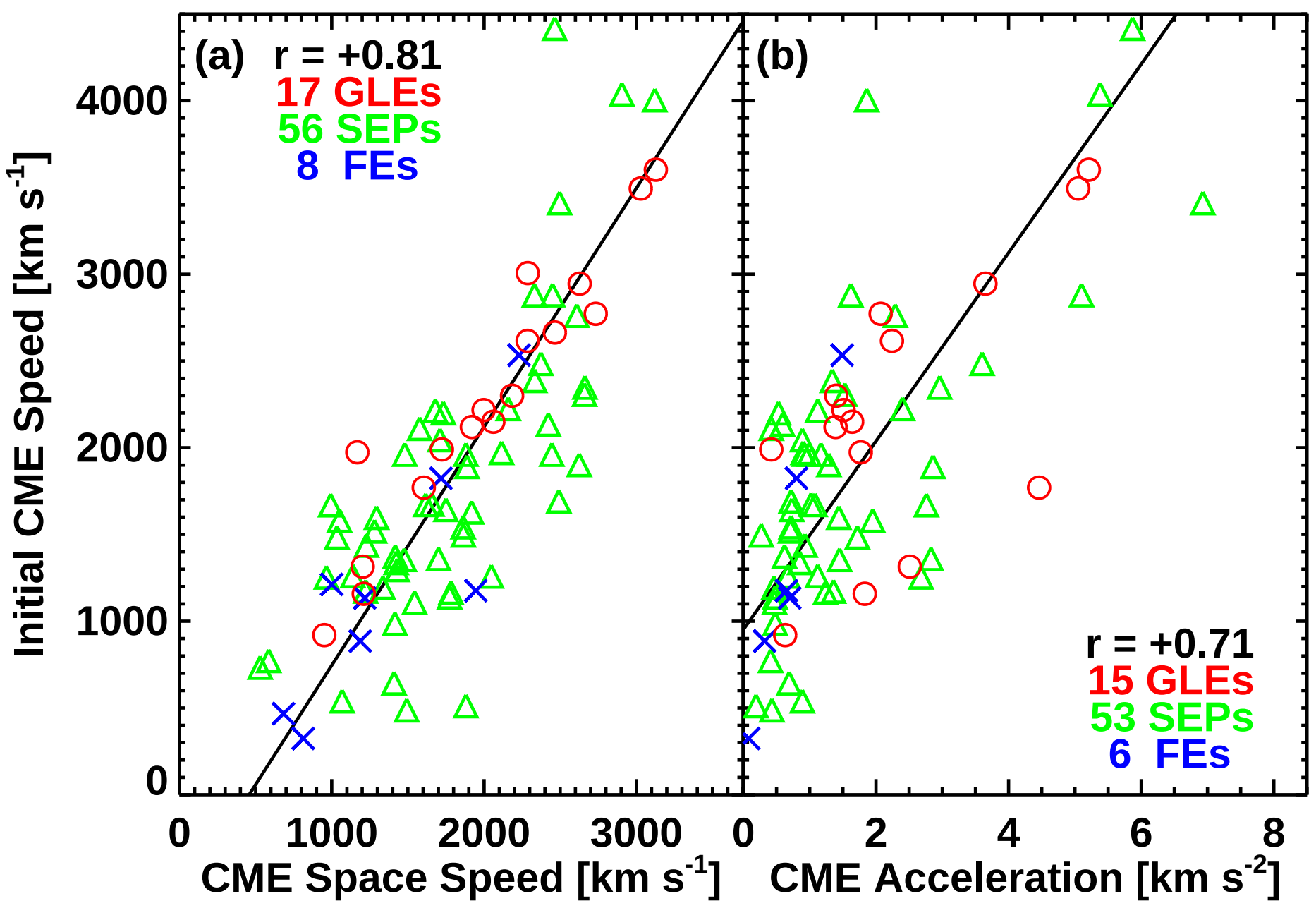

Figure 9. CME initial speed plotted against CME space speed (a) and initial CME acceleration (b). Both plots show high correlation: r =0.81 and 0.71 (the corresponding critical correlations coefficients are $r_c$ = 0.185 and 0.192 at 95% confidence level).

Figure 9 shows the relation between the initial CME speed and space speed. Clearly there is a high correlation (r = 0.81; $r_c$ = 0.185). The CME initial speed is generally higher than the space speed because after attaining a peak speed near the Sun, the speed declines within the coronagraph FOV due to drag. Finally, we also show that the correlation between initial CME speed and the initial acceleration is also high (r = 0.71; $r_c$ = 0.192). This also validates the use of initial CME speed as a proxy to the initial CME acceleration. The advantage of using the initial CME speed is that it can be derived from the first two height-time measurements in the coronagraph FOV, while initial acceleration needs CME space speed and the soft X-ray flare rise time.

### 3.4 CME Kinematics

CMEs associated with large SEP events have been known to have very high average speed (~1500 km $s^{-1}$) in the coronagraph FOV (see e.g., Gopalswamy 2006). In the case of GLE events, CME speeds are even higher, typically ~2000 km $s^{-1}$(see Table 3). In addition, CMEs associated with GLEs accelerate impulsively and attain high speeds very close to the Sun (Gopalswamy et al. 2012a). In this subsection we show that the kinematics of the 2012 July 23 CME are consistent with those of GLE events. The sky-plane speed listed in the SOHO/LASCO CME catalog is ~2003 km $s^{-1}$, which certainly places the July 23 CME into the GLE-producing group.

However, we need to get the acceleration, which requires observations closer to the Sun. We make use of STEREO COR1 and EUVI data for this purpose.

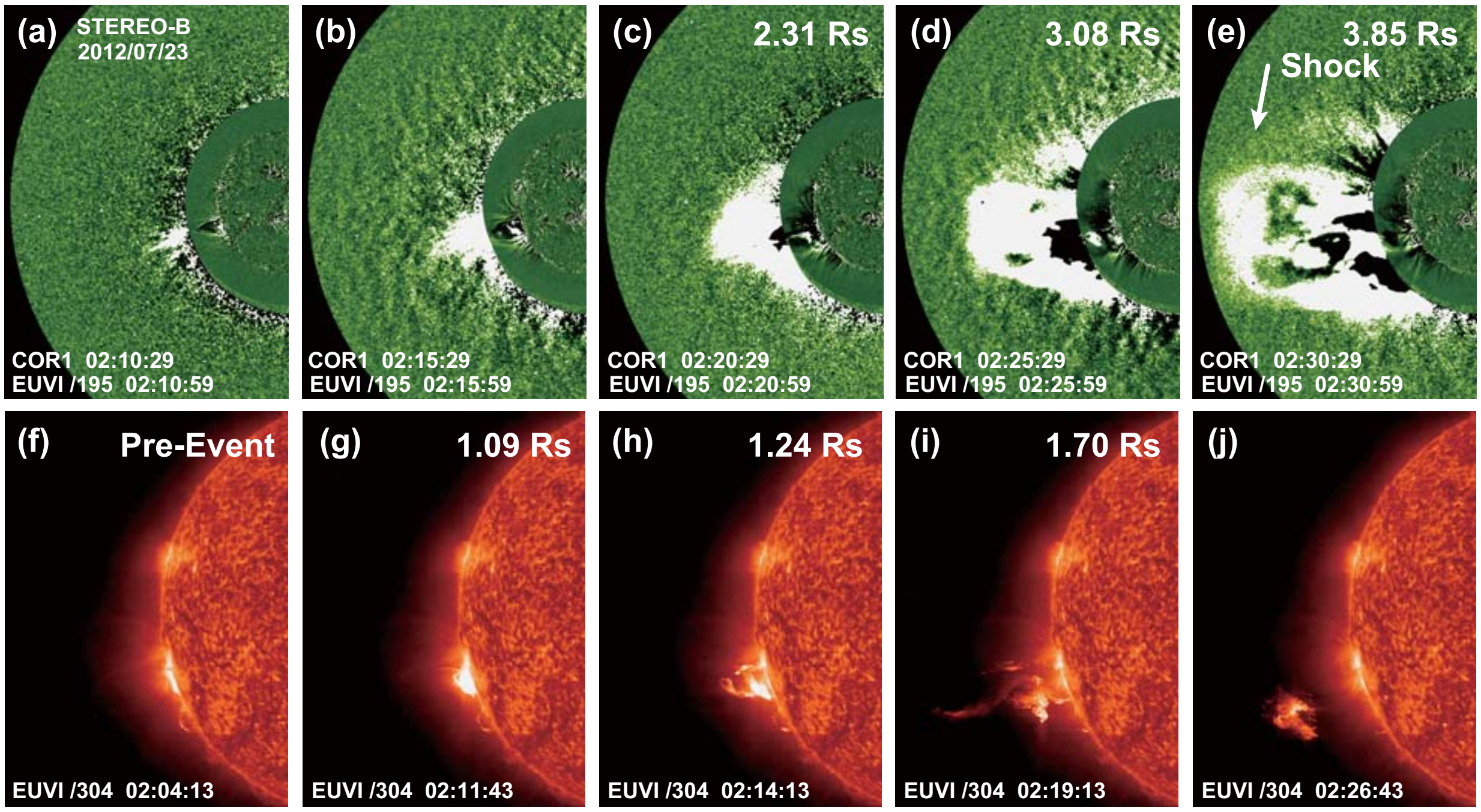

Figure 10 (top) A series of STB/COR1 running difference images showing the onset and early evolution of the 2012 July 23 CME (a-e). In each COR1 image, the nearest EUVI difference image is superposed to show the near-surface activities. The solar limb can be discerned from the EUVI images. The green disk is the image of the COR1occulter with its outer edge at 1.4 Rs. The leading edge heights (from the Sun center) are noted on the images. The height in (b) is unknown because the CME overlapped with the preceding narrow CME in (a). The white-light shock surrounding the CME is marked in (d), but it can also be seen in the previous images. (bottom) The appearance of the prominence core of the CME in the STB/EUVI FOV and its evolution. The heliocentric distances at three instances are noted on the images (g-i). The eruption probably started before 02:04 UT, so it is not seen above the limb in the pre-event image (f). The brightest compact feature above the limb in the EUV difference image in (b) is the same as the brightest feature in (h) and the prominence fragment seen well above the limb in (j). The leading edge of the prominence had left the FOV in (j), so no height is given.

Liu et al. (2014) used the geometric triangulation technique applied to STB and SOHO data and found the July 23 CME speed to be as high as 3050 km $s^{-1}$. Baker et al. (2013) obtained a slightly lower value of ~2500 km $s^{-1}$. Cash et al. (2015) used STEREO beacon data and obtained speeds in the range 1453 to 2985 km $s^{-1}$. They also obtained the CME latitude of 0º and a longitude of ~135º. Applying the GCS model to the science data, Gopalswamy et al (2014a) had obtained a peak leading edge speed of the CME as ~2600 km $s^{-1}$, with a similar shock speed (2580 km $s^{-1}$) obtained by Temmer and Nitta (2015). The latitude and longitude of the flux rope in these works were also similar to those in Cash et al. (2015).

Since we cannot rule out the possibility that the two eruptive filaments are two strands of the same filament, we consider the event to be a single CME. Liu et al. (2014) considered them as independent events. However, there was a narrow and slow preceding CME (see Fig. 10a) that was overtaken by the primary CME in question. We think the preceding CME was not observed beyond ~02:15 UT. For the present purposes, we consider the kinematics of the fast CME as observed by STEREO and SOHO. We also confirmed the prominence evolution using EUV data from the Solar Dynamics Observatory's (SDO) Atmospheric Imaging Assembly (AIA, Lemen et al. 2012), and the SWAP (Sun Watcher using Active Pixel System detector and Image Processing) telescope (Seaton et al. 2013).

In STB (located at E115) view, the CME was only ~14° behind the east limb, so the height-time measurements are close to the true values. The eruption was a disk event (S17W20) in STA view because the spacecraft was located at W121 from Earth. STA/EUVI observed an EUV wave and post-eruption arcade. STA/EUVI observations also show an earlier brightening at 01:40 UT, which was accompanied by a weak EUV wave and dimming (related to the narrow CME observed in STB view as mentioned in the previous paragraph). The main eruption happened around 02:10 UT, which was accompanied by a large-scale EUV wave and post-eruption arcade. In STB view, the earlier eruption resulted in a narrow, jet-like CME that lasted until 02:10 UT (Fig. 10a) and was overtaken by the main eruption at 02:15 UT (Fig. 10b). The interaction made it difficult to determine the leading edge of the CME in this frame. Fig. 10(b-e) show the evolution of the CME in STB/COR1 images. In 10(b), the leading edge of the CME is not clear because it overlapped with the preceding CME, so the heights were measured from the images in Fig. 10(c-e). Fortunately, there were three 304 Å STB/EUVI images that show the evolution of the prominence core (see Fig. 10 (g-i)). The leading edge of the prominence already left the FOV in Fig. 10(j). From the leading-edge heights of the eruptive prominence, we obtained an acceleration of 1.65 km $s^{-2}$, which is in the range of accelerations found in CMEs associated with GLE events (Gopalswamy et al. 2012a). Correcting for projection effects, the acceleration becomes ~1.70 km $s^{-2}$. Using this acceleration, and the measured height of 2.31 Rs at 02:20 UT, we estimate the leading edge height to be ~1.48 Rs at 02:15 UT. This height corresponds to the inner edge of the COR1 FOV, confirming that the leading edge at 02:15 UT is likely to be due to the preceding narrow CME. We were also able to confirm the early CME and prominence heights from the SWAP images at 174 Å (not shown). The leading edge of the prominence at 02:19 UT was measured as 1.30 Rs in SWAP FOV, whereas the actual height was 1.70 Rs from STB/EUVI (see Fig. 10i). At 02:19 UT, the CME leading was at 1.60 Rs in SWAP FOV. Comparing the prominence heights in SWAP and STB/EUVI FOVs, we see that the 02:19 UT CME height of 1.6 Rs implies a true height of ~2.09 Rs, which is in agreement with the heights at 02:15 UT and 02:20 UT.

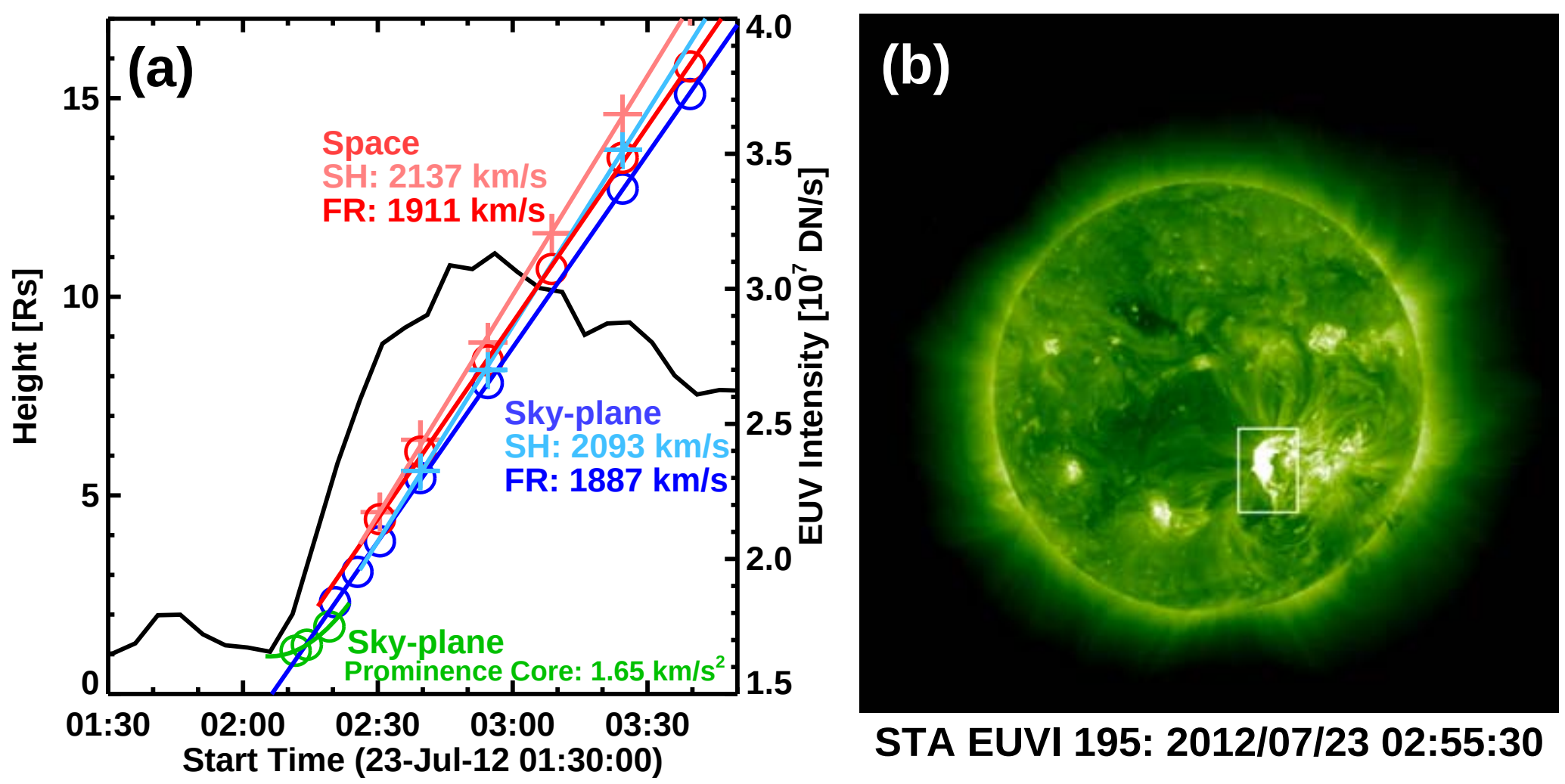

Figure 11. (a) Height-time plot of the CME and its leading shock. The measurements were made at the nose of the flux rope. The blue circles are sky-plane measurements of the flux rope made from STB/COR1 and COR2 images, while the red circles are from the fitted 3D flux rope. The blue and red crosses represent the height of the shock nose in the sky plane and in 3D, respectively. The straight lines are linear fits through the four sets of data points. The sky-plane and 3D speeds are noted near the curves. The three green data points are the leading edge heights of the prominence core observed in three STB/EUVI 304 Å frames before it crossed the EUVI FOV. The three points give an acceleration of ~1.65 km $s^{-2}$ in the sky plane. Correcting for projection effects, the acceleration becomes 1.70 km $s^{-2}$. The EUV intensity of the active region from STA/EUVI in a box of 168×240 pixels shown in (b) is plotted as a function of time. The preceding small flare at 01:40 UT and the large flare associated with the CME in question can be seen. The large tenuous region immediately to the east of the eruption region is an indication that the solar wind behind the CME was likely to be of low density.

Figure 11 shows the height-time measurements from STB COR1 and COR2 for the CME flux rope and the leading shock. We also obtained the three-dimensional height-time history of the fitted flux rope as described in Thernisien et al. (2011). Our measurements are very similar to the ones reported earlier: Gopalswamy et al. (2014a); Temmer and Nitta (2015) because the same technique was used. The peak speeds obtained by these works were similar. Here we consider average speeds in the coronagraph FOV rather than the speed variation because previous works on GLEs used the average space speed in the coronagraph FOV as the representative speed. The sky-plane speed of the flux rope and the shock were 1887 and 2093 km $s^{-1}$, respectively obtained by a linear fit to the height-time data points. A simple geometric deprojection (the active region was ~14° behind the limb in STB view), increases these values to 1945 km $s^{-1}$ and 2157 km $s^{-1}$. The deprojected values agree quite well with the speeds obtained from fitted the flux rope (1911 km $s^{-1}$) and the shock (2137 km $s^{-1}$). A quadratic fit to the height-time data points indicates that the CME was already slowing down at an average rate of ~9 m $s^{-2}$. Such deceleration values are typical of fast CMEs due to the aerodynamic drag of the ambient medium. We also cross-checked our shock speed by measuring the expansion speed of the CME from STA COR2 images and deriving the radial speed using the empirical relationship obtained by Gopalswamy et al. (2009b). This relation connects the radial speed ($V_{rad}$) with the expansion speed ($V_{exp}$), and the angular half width (w) of the CME: $V_{rad} = ½ (1+\cot w)V_{exp}$. From the STA halo CME images, we measured the flux rope speed in the sky plane at several position angles and obtained an average speed of 1186 km $s^{-1}$, which is ½ $V_{exp}$. From STB COR2 image, we measured the flux rope edges at various times. When the edges reached position angles 50° and 170°, the width remained roughly constant ~120°, and thus w=60°. Such values were also present among the widths estimated by Cash et al. (2015). With w = 60°, we get $V_{rad}$ = 1870 km $s^{-1}$, consistent with STB measurements (1945 km $s^{-1}$) and the flux-rope fitting method (1911 km $s^{-1}$). The sky-plane speed of the shock was also measured at the same position angles as the flux rope yielding an average shock speed of ~1322 km $s^{-1}$. This corresponds to a shock expansion speed (the rate at which the diameter of the halo CME increases) of ~2644 km $s^{-1}$.

For a consistency check of the initial acceleration, we use the flare acceleration method (Zhang et al. 2001; Zhang and Dere 2006): assuming that the flare and CME start simultaneously we compute the CME acceleration as the average CME speed in the coronagraph FOV divided by the flare rise time (see also section 3.3.3). The flare rise time is typically obtained from GOES soft X-ray observations. Since we do not have X-ray observations, we use the EUV flare light curve as a proxy to the GOES soft X-ray flux. Fig.11a shows the EUV light curve obtained as the average intensity within a box around the flaring region in STA/EUVI 195 Å images (see Fig. 11b). From the light curve, we estimate the flare rise time as ~25-30 min. Using the average flux rope speed within the STB coronagraph FOV (~1911 km $s^{-1}$), we get an average acceleration in the range ~1.1 to 1.3 km $s^{-2}$, consistent with the acceleration obtained from the leading edge of the prominence core (1.7 km $s^{-2}$). The height-time analysis confirms that the July 23 CME had the speed and initial acceleration consistent with those of typical GLE events. It must be pointed out that our shock speed (2137 km $s^{-1}$) in the coronagraph FOV is ~30% smaller than the speed obtained by Liu et al. (2014). For a spherical expansion, our shock speed can be as high as ~2600 km $s^{-1}$, which is still about 13% below Liu et al. (2014) value, but close to their error bar. The triangulation technique used by Liu et al. (2014) may have a large uncertainty because identification of common features between STB and SOHO is difficult for a CME occurring ~50° behind the limb in SOHO view. We therefore conclude that the shock speed in the July 23 event well exceeded 2000 km $s^{-1}$.

### 3.5 Height of Shock Formation from Type II radio Burst Observations

One of the key requirements for the acceleration of GeV particles by a CME-driven shock is that the shock forms close to the Sun (see also section 3.3.3). Reames (2009a,b) showed that one can readily explain the height of solar particle release in most of the GLE events if a shock formation height of ~1.5 Rs is assumed. Gopalswamy et al. (2012a; 2013b) confirmed this by estimating the height of shock formation as the CME leading-edge height at the onset of metric type II bursts. When routine observations of CMEs became available, it was found that metric type II bursts start when CME leading edges are in the range ~1.2 to 2 Rs (Cliver et al. 2004; Gopalswamy et al. 2005b; 2009c; 2012a,b; 2013b; Mäkelä et al. 2015). In addition to the small shock-formation height, the shock speed has to be sufficiently high for a strong shock. For example, the 2010 June 13 shock formed at a height of 1.19 Rs and produced a metric type II radio burst, but no SEPs (Gopalswamy et al. 2012b). The shock speed was ~600 km $s^{-1}$ near the Sun, but decreased to 320 km $s^{-1}$ in the LASCO FOV. Another event on 2003 November 2 had a similar shock formation height, but the speed was >2000 km $s^{-1}$ and was associated with a GLE event. Thus, the relative increase of the CME speed with respect to the Alfven speed is a key factor deciding the shock strength (Gopalswamy et al. 2001a).

The Culgoora radio spectrograph (Thompson et al. 1996) recorded a metric type II burst in association with the 2012 July 23 CME. The narrow band type II burst drifted from 45 MHz (02:31 UT) to 20 MHz (02:37 UT). For a behind-the-limb event, the emission is likely to be in the harmonic component because the foreground plasma would absorb the fundamental. This means the plasma frequency at the source region when the type II burst started was ~22.5 MHz (electron density ~$6.3\times10^6$ $cm^{-3}$). The dynamic spectra in Fig. 9 that combine observations from the Culgoora radio spectrograph with STEREO/WAVES and Wind/WAVES show that the metric type II burst continues as the harmonic component (H1) in the decameter-hectometric (DH) domain. The fundamental component (F1) is seen from 16 MHz (the highest frequency of the SWAVES receiver) down to ~1 MHz. The F1 component was observed only at 02:54 UT by Wind/WAVES because of occultation. Another intense pair F2-H2 also started around this time at lower frequencies and will be discussed in more detail later in the subsection.

There was an intense group of type III bursts observed both by STEREO and Wind. In STA, the first type III occurred at 01:44 UT. This one and the second type III burst at 01:55 UT seem to correspond to the first smaller eruption (see the EUV light curve in Fig. 10a). The complex type III burst starting at 02:03 UT and ending at 02:32 UT is from the main eruption in question. The onset of the type III burst also identifies the onset of the eruption that resulted in the large CME, consistent with the EUVI images shown in Fig. 12. All the type III bursts were also observed by Wind/WAVES, but only below ~2 MHz because of occultation. Only two of the intense bursts within the complex group were observed by Wind/WAVES at higher frequencies; these are probably harmonic type III bursts. The complex type III burst also probably masked the onset of the F2-H2 pair if it existed before 02:30 UT.

According to Gopalswamy et al. (2013b) the height of shock formation corresponding to22.5 MHz can be estimated as 2.0 Rs, if the type II burst was originating radially above the source region. Since the source region was ~50º behind the limb, the starting frequency of the type II burst is likely to be much higher than 45 MHz (or the fundamental component at frequencies higher than 22.5 MHz). The observed starting frequency of 45 MHz therefore corresponds to the highest frequency that can pass through the foreground corona.

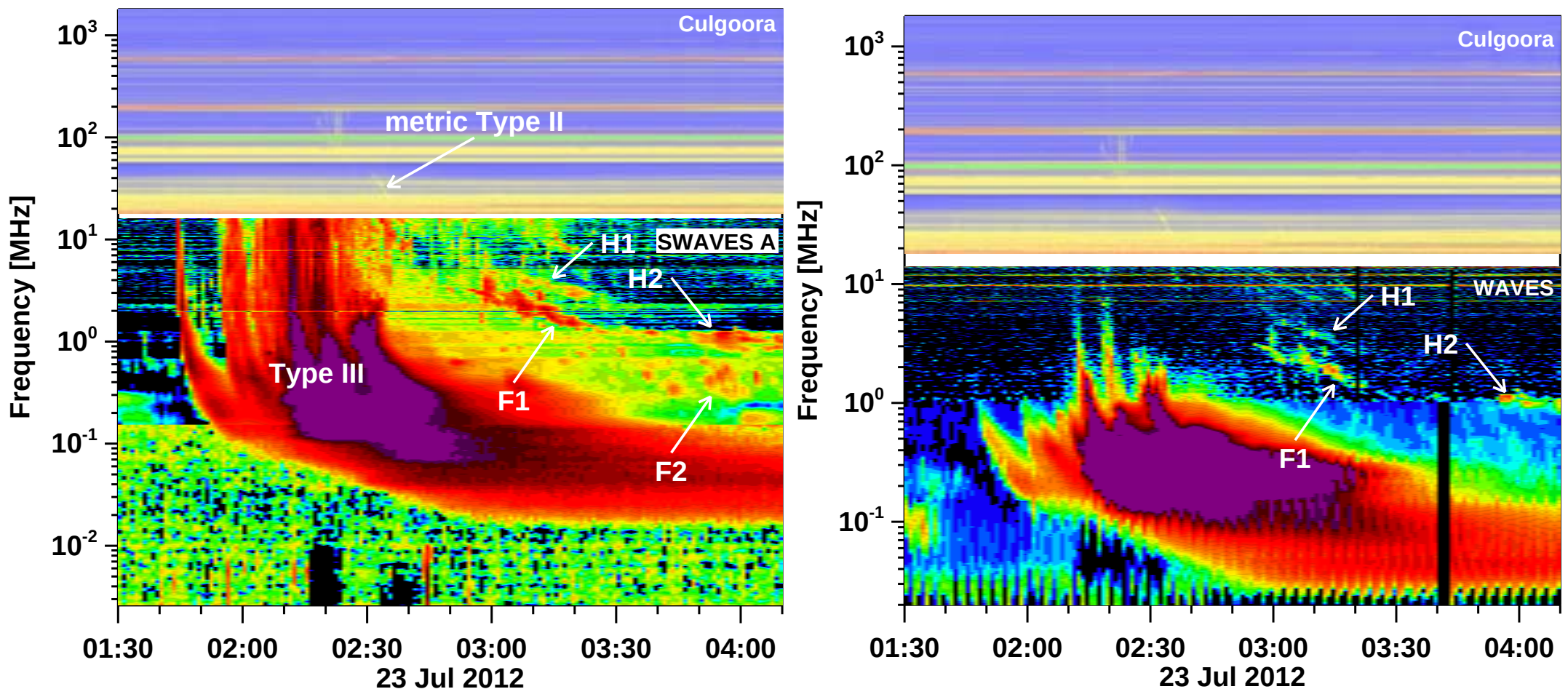

Figure 12. Composite dynamic spectra of the 2012 July 23 backside event that combines ground based observations from the Culgoora radio spectrograph with space observations from STEREO/WAVES Ahead (left) and Wind/WAVES (right). The metric type II burst drifts from 45 MHz down to 20 MHz and appears to continue as the harmonic component H1 in the decameter-hectometric (DH) domain; the fundamental component F1 can be seen up to the highest SWAVES frequency (~16 MHz). F1 and H1 are also observed by Wind/WAVES but the F1 component is occulted above 2 MHz. The intense group of type III bursts is marked in the STA/WAVES dynamic spectrum. Most of the type III bursts in Wind/WAVES are also occulted above 1-2 MHz, similar to F1.

From the height-time plot in Fig. 11a, we see that the CME nose was already at a heliocentric distance of ~5 Rs at the time of the metric type II burst. The local plasma frequency at 5 Rs is expected to be much lower than 22.5 MHz, more like ~0.6 MHz (see Gopalswamy et al. 2013b). Indeed there was an intense type II burst in progress at lower frequencies with fundamental harmonic structure (F2-H2) observed by STA (see Fig. 12). Such low frequency emissions cannot pass through the foreground plasma to reach Wind/WAVES. The metric type II emission has to come from a lower heliocentric distance, which is possible at the flanks of the shock between 2 and 5 Rs. The bandwidth of the metric type II burst is extremely narrow suggesting that radio emission came from a small section of the shock located on its earthward flank.

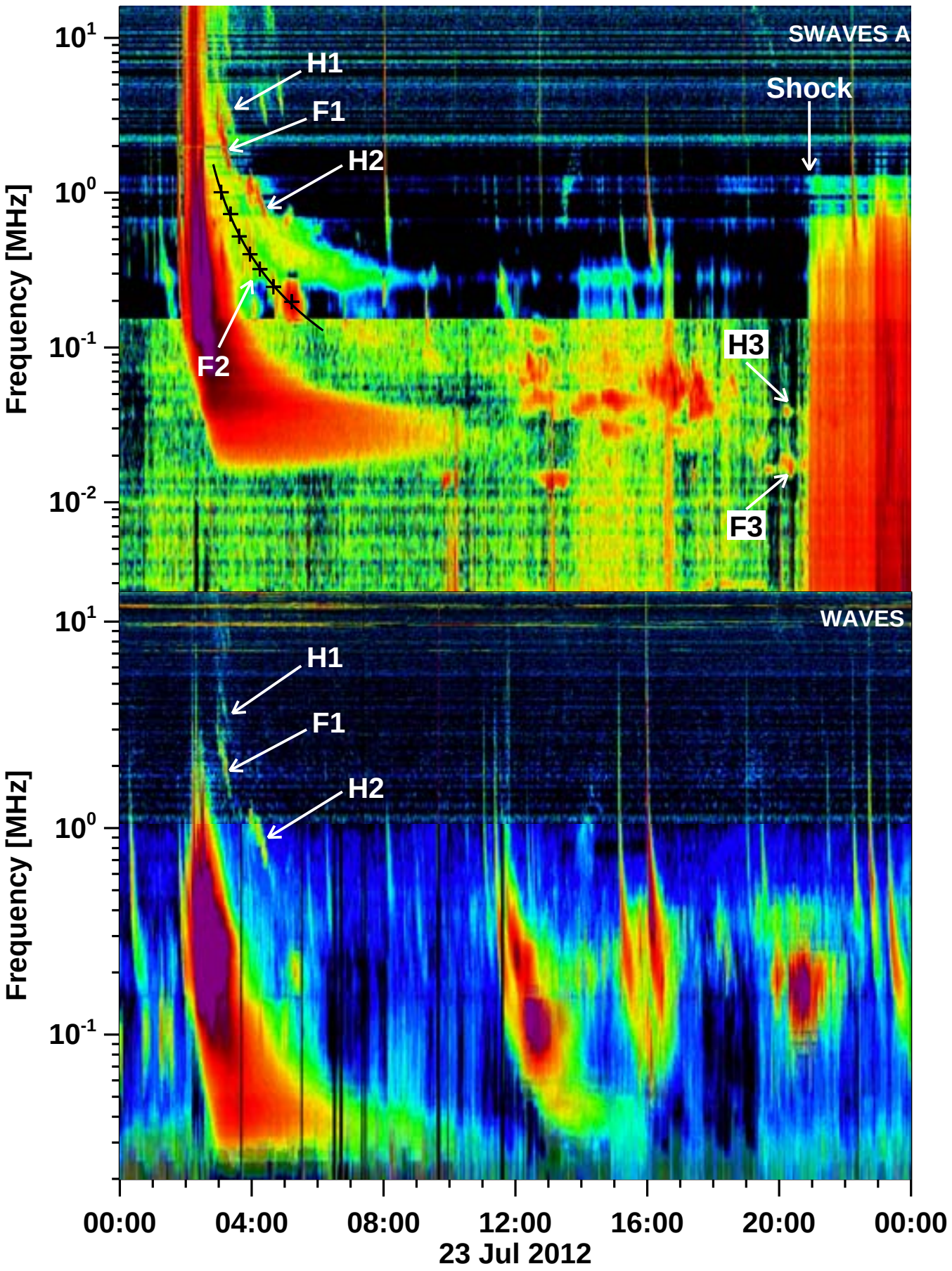

Figure 13. Dynamic spectra of the 2012 July 23 event from STA/WAVES (top) and Wind/WAVES (bottom). F2 and H2 are the fundamental and harmonic components of the intense type II burst starting below 2 MHz. The burst becomes patchy below 0.2 MHz, but continues all the way until the shock arrival at 1 AU (the local type II burst also has F-H structure marked F3 and H3). In the Wind/WAVES spectrum, only the brightest part of the harmonic H1 and a brief fragment of the fundamental F1 are observed. Also, the type III bursts and F1 are occulted above ~2 MHz. The two type III bursts observed up to the upper frequency limit of Wind/WAVES are likely to be harmonic components.

The type II burst with intense F-H components started below 2 MHz (harmonic) before the end of the flank component, which had started in the metric domain (see Fig. 13). This component has to be from the nose, where the shock is the strongest. This can be readily confirmed by computing the shock speed ($V_s$) from the drift rate (*df/dt*) of the fundamental component of the intense type II burst using the following relation:

$$V_s = 2L(1/f)(df/dt), \quad (1)$$

where $f$ is the reference frequency around which the drift rate is measured and $L$ is the local density scale height,

$$L = |(1/n)(dn/dr)|^{-1} \quad (2)$$

given by the distribution of coronal density ($n$) as a function of the heliocentric distance $r$. If $n \sim r^{-\alpha}$, one gets a simple relation,

$$L = r/\alpha. \quad (3)$$

Figure 13 shows that F2 and H2 of the type II burst are discernible until about 10:00 UT. After this time, there were several blobs of bright emission, but the F-H structure is not clear. However, when the shock arrived at STA, one can see local type II emission with the fundamental and harmonic structures marked as F3 (17 kHz) and H3 (34 kHz). It is highly likely that they are the continuation of F2 and H2. We were able to determine the drift rate from both F and H components. We selected several frequencies ($f$) along F2 and noted the corresponding times (the black line in Fig. 12). We converted the times to distances from the shock height-time measurements. Plotting $f$ as a function of $r$, we found a power law form $f \sim r^{-1.5}$. Since $f$ is the local plasma frequency we see that $n \sim r^{-3}$. In other words, we need $\alpha = 3$ to get the shock speed from the dynamic spectrum to match the shock speed from coronagraph images. The derived $\alpha$ is an intermediate value between $\alpha = 4$ in the DH domain and $\alpha = 2$ in the kilometric domain.

F1 is the fundamental part of the type II burst that continued from the metric wavelengths. Considering the segment of F1 from 02:56 – 03:19 UT drifting from 3.21-1.49 MHz gives a drift rate of $1.25\times10^{-3}$ MHz $s^{-1}$. At a reference frequency of $f$=2.35 MHz at ~03:07 UT, the height-time history in Fig. 11 places the shock nose at $r$ = 11.4 Rs. However, F2-H2 at lower frequencies was in progress around this this time from the nose part. Therefore, F1 must be occurring at heights smaller than 11.4 Rs, which means the flank. We can estimate the height assuming that the density variation with height is similar above the nose and flanks of the shock and $f \sim r^{-2}$ (corresponding to $\alpha = 4$) as 7.4 Rs. The speed at this height from equation (1) becomes 1377 km $s^{-1}$, which is quite reasonable for the shock flank. The flank speed is ~1200 km $s^{-1}$ if $f \sim r^{-1.5}$ (appropriate to the density variation for F2-H2) is used. Recall that H1 is the continuation of the metric type II burst (see Fig. 12). The harmonic component drifts from 45 to 20 MHz. Equivalently, the fundamental drifts from ~22.5 MHz to 10 MHz in 6 min, yielding a drift rate of ~0.035 $MHzs^{-1}$. For $\alpha = 6$ in the corona (Gopalswamy 2011) and a reference frequency of 16 MHz, we get a speed of ~1030 km $s^{-1}$ for a heliocentric distance of 2 Rs from equations (1) and (2). This speed is consistent with the flank speed derived from STA/WAVES data for F1.

Although the metric type II burst indicates that a shock existed at 02:31 UT, it does not tell us when and where the shock formed because of the occultation. We can estimate the shock formation height from the CME speed early in the event. From Fig. 11, we see that the CME had attained a speed of ~1900 km $s^{-1}$ at 02:20 UT. The CME first appeared in the STB/COR1 FOV at 02:15 UT. Using the initial acceleration we derived from STB/EUVI images of the prominence core (1.70 km $s^{-2}$), we estimate the CME speed at 02:15 UT to be ~1400 km $s^{-1}$. Recall that the CME leading edge was at ~1.48 Rs at 02:15 UT. The typical fast mode speed around 1.5 Rs is ~400 km $s^{-1}$ (Gopalswamy et al. 2001a; Mann et al. 2003), so a CME speed of ~1400 km $s^{-1}$ indicates a magnetosonic Mach number of 3.5. Thus we conclude that the shock formation height should be at 1.48 Rs or smaller. Zhu et al. (2016) performed a velocity dispersion analysis of energetic particles during the July 23 event and obtained a particle release time of ~02:26 UT at the Sun. It is well known that it takes typically about 10 minutes for the particles to be released after the shock formation (Reames 2009a,b; Mewaldt et al. 2012; Gopalswamy et al. 2012a; 2013a). Therefore, a shock formation around 02:15 UT is consistent with the solar particle release time of ~02:26 UT. The shock height at 02:26 UT is ~3.5 Rs, which is also typical of particle release heights in high-energy SEP events (Gopalswamy et al. 2012a, their fig.16).

In summary, both coronagraph and radio spectrograph observations suggest that the shock speed was >2000 km $s^{-1}$ early in the event. Such high speeds are typical of GLE events when well connected to Earth (Gopalswamy et al. 2012a). Thus a strong shock is implied, which continued to produce type II burst all the way to 1 AU until it arrived at STA as seen in Fig. 13. Type II bursts with emission components from the metric (m) to kilometric (km) wavelengths are indicative of strong shocks throughout the inner heliosphere that cause intense SEP events (Gopalswamy et al. 2005b).

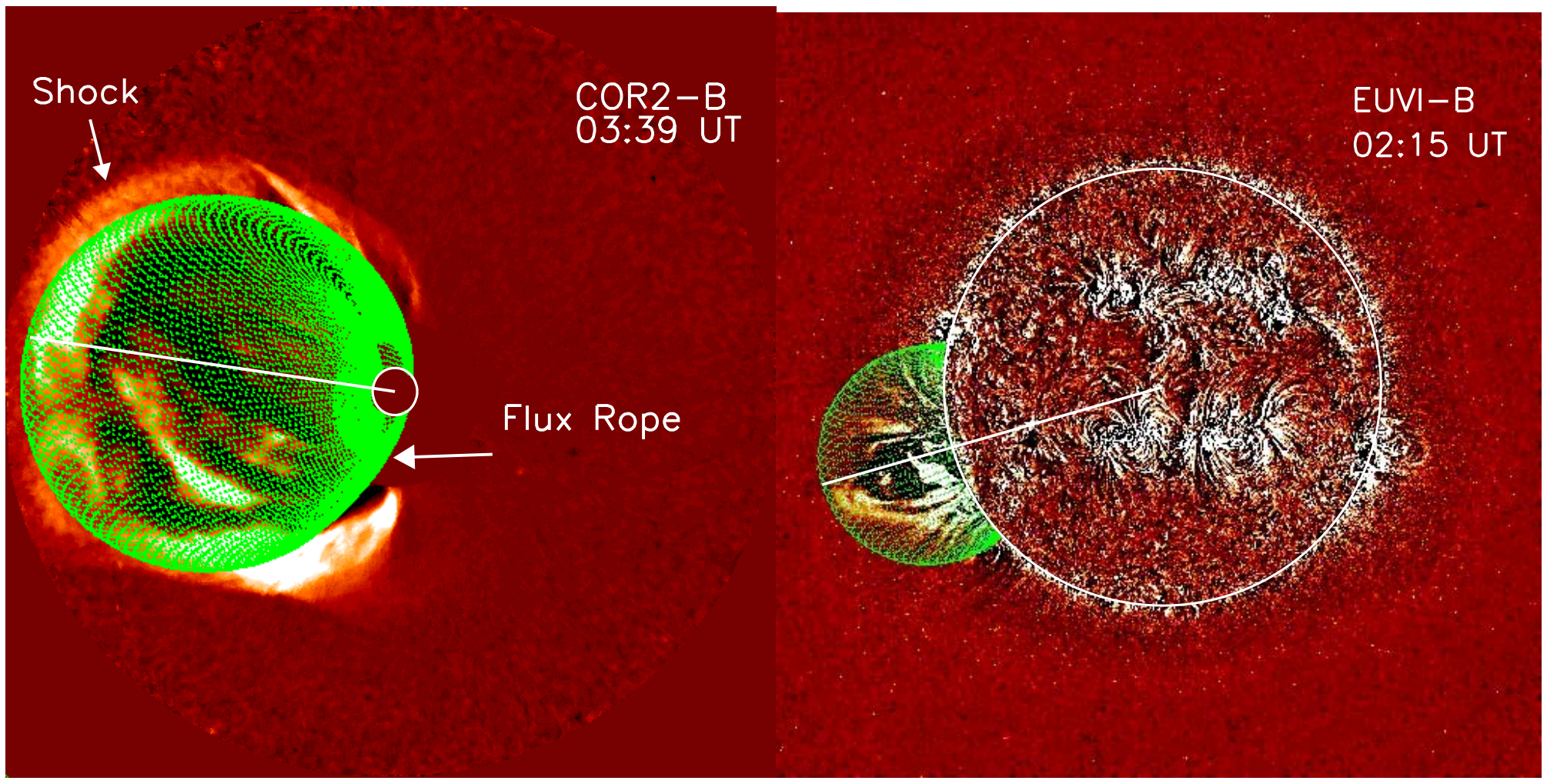

Figure 14. The 2012 July 23 CME and shock with the flux rope fit (green mesh) superposed: (left) COR2 CME at 03:39 UT and EUVI CME at 02:15 UT. The white line connects the Sun center to the nose of the flux rope. Note that the flux rope deflected to the north so that the effective latitude of the flux rope is at the ecliptic.

**4. Discussion**

The main results of this paper are (1) the backside eruption of 2012 July 23 qualifies as a historical extreme event from the energetic particles perspective, and (2) there is a hierarchical relationship between initial CME acceleration and the fluence spectral index. This paper has drawn upon disparate data sets to illustrate this point. We considered eight characteristics that support the likelihood that the 2012 July 23 back side CME was indeed an extreme event:  high SEP intensity, hard TOM spectrum, hard fluence spectrum, high CME speed, rapid initial acceleration, small shock formation height, m-km type II burst, and equatorward deflection of the CME.

### 4.1 Comparison with the 1991 March 23 SEP event

We compare the 2012 July 23 event with the largest front-side SEP event since 1976: the 1991 March 23 which had an ESP peak of 43,000 pfu and an SEP peak of 4570 pfu in the >10 MeV channel. The corresponding values for the 2012 July 23 event were estimated as 43,700 pfu and ~4650 pfu. Interestingly, the 1991 March 23 event was not a GLE (Shea and Smart 1993), so why should the 2012 July 23 event have accelerated particles to GeV energies? The lack of GeV particles in the 1991 March 23 event is likely due to the poor connectivity to Earth both in longitude (E28) and in latitude (S26). Accounting for the B0 angle (~-7°), the ecliptic distance of the source becomes ~19°, which is larger than the typical ecliptic distance of~13° for GLE events (Gopalswamy et al. 2013a). We examined the coronal hole map corresponding to the Carrington rotation period including the 1991 March 23 event. We did not see any large coronal hole that might have deflected the CME closer to the equator. On the other hand, the 2012 CME was at western longitudes (W20) in STA view and the effective source location was right at the ecliptic. This is illustrated in Fig. 14, where we have shown the fitted flux rope at two instances: 02:15 UT when the leading edge was close to the limb in STB view and at 03:39 UT when the flux rope was in the STB/COR2 FOV. Clearly, the line connecting the Sun center to the flux rope leading edge deflected equatorward by ~20°. The flare location was at S17W141, but the location of the fitted flux rope was N05W135, which becomes N00W135 when corrected for the B0 angle. In summary, the 2012 July 23 event was better connected magnetically to the observer both in latitude and longitude than the 1991 March 23 event was. This does not rule out the possibility that the 1991 March 23 event did accelerate particles to GeV energies – the most likely scenario is that the nose (where GeV particles are accelerated) was not connected to Earth.

### 4.2 Fluence Spectra

Mewaldt et al. (2012) considered three different functional forms to fit the proton fluence spectrum of cycle-23 GLE events: Bessel function (McGuire and von Rosenvinge, 1984), double power law (Band et al. 1993), and power law with an exponential rollover (Ellison and Ramaty, 1985). Here we use a simple power law $dN/dE = AE^{-\gamma}$ (A is a constant and $\gamma$ is the spectral index) to fit the proton fluence in the energy range 10-100 MeV. This choice was mainly motivated by the fact that it nicely fits the fluence spectra of FE SEP events (Gopalswamy et al. 2015b) and other SEP events associated with CMEs with positive acceleration profile in the coronagraph field of view (<30 Rs) irrespective of their source region. Therefore, use of the 10-100 MeV spectral index in characterizing the SEP events, even though the spectral behavior may be different outside this range. For example, the fluence may level off at energies below 10 MeV and there may have a rollover beyond 100 MeV. In some cases, there may be a rollover even in the 10-100 MeV range. In spite of this limitation, we find that the 10-100 MeV fluence spectral index $\gamma$ organizes the SEP events into three groups: FE SEP events, regular SEP events, and GLE events with respectively high, intermediate, and low values of γ. When data points at energies higher than 100 MeV were included in computing the power laws index, it was found that the fluence spectral index was slightly larger (by ~15%), indicating a rollover in many cases. When the rollover is not in the 10-100 MeV range, Mewaldt et al.'s (2012) γ2 (spectral index above the spectral break) was close to our γ. For example, γ2 for the 2005 January 20 and 2006 December 13 events were 2.14 and 2.42, respectively compared to 2.13 and 2.07 in the 10-100 MeV range (see Fig. 6b). The higher value of γ2 results because of a significant rollover at higher energies in the 2006 December 13 event, whereas there is none in the 2005 January 20 event. In fact, their γ2 also ordered the GLE and non-GLE events: softer spectra in non-GLE events (<γ2> = 4.34) compared to GLE events (<γ2> = 3.18) (their Fig. 7; they did not consider the FE SEP group separately from the non-GLE SEP events).

#### 4.2.1 Comparison with Some Historical Events

It is also worth comparing the >30 MeV fluence, which has been extensively used to assess the significance of historical events such as the Carrington event. Cliver & Dietrich (2013) made an estimate of the omnidirectional integral fluence of protons $1.1\times10^{10}$ pr $cm^{-2}$ for the Carrington event. They also estimated the >30 MeV fluence of several modern SEP events (July 1959, November 1960, and August 1972) to be in the range $5\text{-}7\times10^{9}$ pr $cm^{-2}$. These values are comparable to that in the July 23 event ($2.1\times10^{9}$ pr $cm^{-2}$) estimated from the STA/HET data assuming that particles beyond 1000 MeV did not contribute to the fluence (see also Mewaldt et al. 2013). The July 23 fluence is only about a factor 5 smaller than that estimated for the Carrington event. The 1956 February 23 GLE event from a longitude similar to that of the July 23 event had only a >30 MeV fluence of $1.0\times10^{9}$ pr $cm^{-2}$, which is a factor of 2 smaller than the July 23 fluence. Cliver & Dietrich (2013) also derived the absolute integrated-fluence spectra of the 1972 August 4 and 1956 February 23 GLE events. In Figure 15, we have plotted the data points from figure 13 of Cliver and Dietrich (2013); we obtain an integrated-fluence spectral index of 1.80 and 1.03 for the 1972 August and 1956 February events, respectively using the two data points in the 10-100 MeV range. The corresponding index for the July 23 event depends on the cutoff energy used beyond the 100 MeV. For cutoffs in the range 150 to 1000 MeV, we get an integrated-fluence spectral index of 1.82 to 0.69. In Figure 15, we see that the spectra with 1000 MeV and 500 MV cutoffs are not smoothly decreasing, indicating that the > 100 MeV point is too high. The average index from the remaining three cases is 1.46, which is midway between the August 1972 and February 1956 events. In the worst case, the July 23 event has the same integral fluence spectral index as the August 1972 event. Thus, we conclude that the July 23 event is similar to the 1972 August and 1956 February in terms of the integrated-fluence spectrum; this comparison further supports the possibility that the July 23 event accelerated GeV particles and was an extreme particle event.

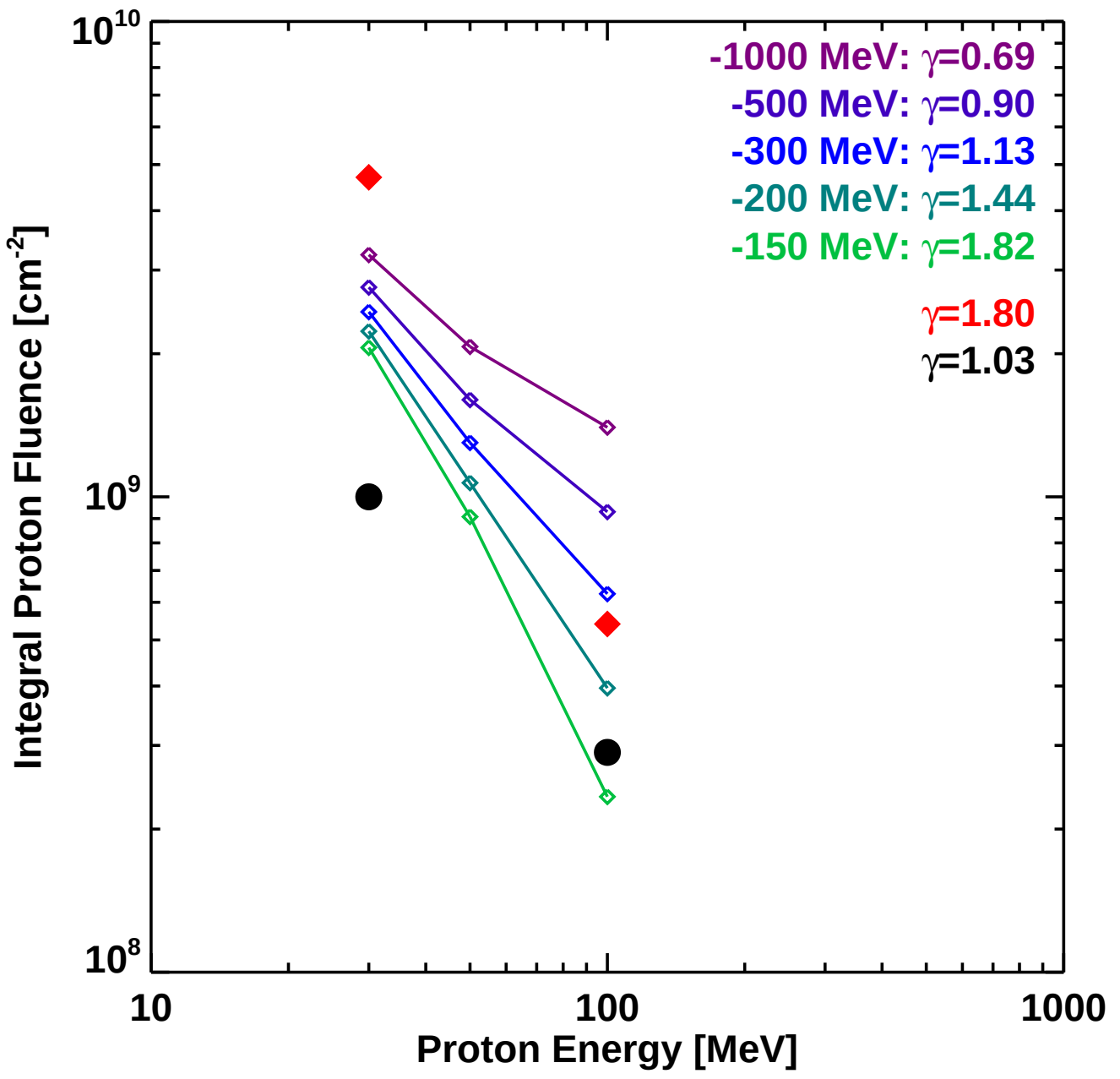

Figure 15. The integrated fluence spectrum of the July 23 event compared with those of the 1972 August 4 (red diamonds) and 1956 February 23 (filled black circles) events. The spectra for the July 23 event were computed at >30 MeV, >50 MeV, and >100 MeV using STA/HET data. The five different spectra were obtained assuming cutoff energies ranging from 150 MeV to 1000 MeV. The corresponding spectral indices are shown on the plot.

### 4.3 CME Kinematics and ICME Observations

We derived the average shock speed in the coronagraph FOV as 2137 km $s^{-1}$ with an initial acceleration of ~1.70 km $s^{-2}$. These values are quite similar to the corresponding values obtained for the two cycle-24 GLE events: 1997 km $s^{-1}$ and 1.77 km $s^{-2}$ for the 2012 May 17 event (Gopalswamy et al. 2013a) and 1960 km $s^{-1}$ and 2.08 km $s^{-2}$ for the 2014 January 06 event (Thakur et al. 2014). The acceleration we obtained is also consistent with the peak acceleration of 2.03 km $s^{-2}$ obtained by Temmer and Nitta (2015), but ours is an average acceleration.

The shock speeds reported in the literature have been controversial both near the Sun and at 1 au. We already discussed that the average CME and shock speed in this work is smaller than the values estimated by Liu et al. (2014) using the geometric triangulation method. Note that our values are from direct measurements from STB data with no modeling. Flux-rope fitting to the CME images obtained from the three views (SOHO, STEREO) also agreed with the STB direct measurements (see also Gopalswamy et al. 2014a; Temmer and Nitta 2015). We also note that the transit speed of the shock (~2200 km $s^{-1}$) is comparable to the coronagraph shock speed (~2137 km $s^{-1}$), suggesting that there was no deceleration of the shock between the Sun and Earth as pointed out by Liu et al. (2014). From the in-situ data plotted in Liu et al. (2014), we see that ICME1 and ICME2 had leading edge speeds of 1680 km $s^{-1}$ and 1560 km $s^{-1}$, respectively. If we consider that ICME2 is the counterpart of the near-Sun flux rope and that ICME1 is the shock sheath (preceding CMEs can be part of shock sheaths), we see a deceleration of the flux rope from 1911 km $s^{-1}$ at the Sun at the rate of 5.3 m $s^{-2}$. The empirical acceleration model of Gopalswamy et al. (2001b) connects the interplanetary acceleration $a$ to the CME speed $V$ in the coronagraph FOV: $a = 0.0054$ ($V$-406) m $s^{-2}$. Substituting $V$=1911 km $s^{-1}$, we get $a$ = 8.1 m $s^{-2}$. Clearly, the observed deceleration is ~34.5% lower than the average value. One possibility is that the solar wind speed was higher than 406 km $s^{-1}$. For example if we use a solar wind speed of 450 km $s^{-1}$, the deceleration becomes 8.1 m $s^{-2}$. Temmer and Nitta (2015) found that for the Drag Based Model (Vršnak et al. 2013) to yield results consistent with observation, the ambient density should not exceed 1-2 $cm^{-3}$. This requirement is in contradiction with the observation of the in-situ type II burst F3-H3. The fundamental and harmonic components were at 0.017 and 0.034 MHz, which yield a local plasma density of 3.6 $cm^{-3}$ upstream of the shock. Further investigation is needed to resolve this contradiction. One other peculiarity we would like to point out is the extremely high expansion speed of ICME2. From the leading-edge ($V_L$ = 1560 km $s^{-1}$) and trailing-edge ($V_T$ = 875 km $s^{-1}$) speeds, we get an expansion speed of ~685 km $s^{-1}$, which is more than three times the highest expansion speed observed in cycle-24 magnetic clouds (Gopalswamy et al. 2015d). The expansion speed is ~44% of the leading edge speed, while the typical value is ~6.3%. The central speed ($V_C = ½ (V_L+V_T)$) is also relatively small (~1218 km $s^{-1}$), again indicating a rapid expansion. The high total pressure inside the ICME and a low ambient total pressure can cause such a large expansion. On the other hand, the expansion factor $V_L/V_c$ = 1.28 is within the range of values reported for magnetic clouds in cycle 23 and cycle 24 (Gopalswamy et al. 2015d).

The in-situ shock speed at STA has also been controversial, primarily because the lack of adequate plasma measurements in the vicinity of the shock. Riley et al. (2016) estimated the density around the shock and obtained a shock speed of 3377 km $s^{-1}$. In the strong shock limit, the shock speed was already estimated by Riley et al. (2016) as 2700 km $s^{-1}$, although the shock is a cosmic-ray modified shock, in which the maximum compression ratio need not be 4. One other estimate of the shock speed without using density measurements in the shock vicinity is to use the gas dynamic relation between the shock speed ($V_s$) and the ICME speed ($V_p$):

$$V_s = V_p \, (1+\gamma_a)/2 \qquad (4)$$

where $\gamma_a$ is the adiabatic index. Using $V_p$ = 1560 km $s^{-1}$ and $\gamma_a$ = 5/3, we get $V_s$ = 2080 km $s^{-1}$, consistent with the shock speed near the Sun. If the driver speed is taken as $V_p$ = 1680 km $s^{-1}$, $V_s$ would go up to 2240 km $s^{-1}$. One possibility is to make use of the constraint provided by the local type II radio burst (F3, H3) that the upstream density is ~3.6 $cm^{-3}$ to better constrain the 1-au shock speed.

## 5. Summary and Conclusions

The 2012 July 23 CME from behind the Sun was associated with an interplanetary CME of high magnetic content, a fast shock, and a large SEP event. The SEP event was a circumsolar event observed at both STREO spacecraft, SOHO, GOES, and MESSENGER. The shock transit time was ~18.5 h. Only 15 such extreme fast-transit events have been reported since the Carrington 1859 event (Gopalswamy et al. 2005a). Therefore, the July 23 event adds one more event to this historical list. The CME has been deemed to be of Carrington class based on its magnetic content. In this paper, we have shown that the CME is also extreme from the particle acceleration perspective: not only are the SEP and ESP intensities very high, the fluence spectrum of the event is consistent with SEP events consisting of GeV particles. In order to demonstrate the last point, we determined the 10-100 MeV fluence spectral index of cycle 23 and 24 SEP events and showed that the July 23 event belongs to the GLE group. The >30 MeV integral fluence of this event is similar to that in historical events such as the 1972 August 4 and 1956 February 23 GLE events, strongly supporting the result that the July 23 event was in deed an extreme event from the standpoint of accelerating the highest energy particles. Other conclusions of the paper are:

1. The >10 MeV SEP and ESP intensities of the 2012 July 23 event exceed the corresponding values of all large SEP events detected by GOES since 1976.

2. The 10-100 MeV time of maximum spectrum of the 2012 July 23 SEP event is very small ($\gamma$ ~1.30) consistent with GLE events.

3. The indices of fluence spectra of SEP events in solar cycles 23 and 24 fall into three groups: high values ($\gamma \sim 4.89$), intermediate values ($\gamma \sim 3.83$), and low values ($\gamma \sim 2.68$) belonging to SEP events associated with filament eruptions outside of active regions (FE SEP events), regular SEP events, and GLE events, respectively. The 2012 July 23 event had $\gamma \sim 2.28$, clearly belonging to the GLE group.

4. There is a hierarchical relationship between initial CME acceleration and the power law index of the fluence spectrum: FE SEP events have the softest spectra and lowest initial acceleration; GLE SEP events have the hardest spectra and the largest initial acceleration; the regular SEP events have intermediate spectral indices and acceleration.

5. Even though there are some discrepancies in the literature on the speed of the CME-driven shock near the Sun for the July 23 event, we showed that the speed certainly exceeds 2000 km $s^{-1}$, which is the average speed of GLE-producing CMEs. The ICME speed was also very high and appropriately high shock speed, although the actual values are still not established due to lack of plasma measurements at the time of the shock.

6. The initial acceleration of the July 23 CME was ~1.7 km $s^{-2}$, also typical of GLE-producing CMEs in cycles 23 and 24.

7. The fluence spectral index has a significant correlation with the initial CME speed and initial CME acceleration, suggesting that the initial CME speed is a good proxy for the initial acceleration.

8. The fluence spectral index has no dependence on the source longitude when western events are considered.
9. We estimated the shock formation height to be at least 1.48 Rs from metric type II burst observations and the attainment of super-Alfvenic speeds early in the event. Such low shock formation heights is typical of GLE events.
10. We confirm that the interplanetary deceleration of the CME in the July 23 event was smaller than usual: ~35% below the average deceleration. This may be due to a faster solar wind that followed the CME.
11. The July 23 event was accompanied by a type II burst with emission components from metric to kilometric wavelengths. This is typical of all GLE events. Both flank and nose emissions were observed and the speeds derived from the radio dynamic spectra are consistent with the shock speeds at the nose and flanks.
12. The type II burst with fundamental and harmonic components was also observed when the shock arrived at STEREO Ahead. This observation implies an upstream plasma density of ~3.6 $cm^{-3}$, which is a useful constraint that models need to address.

We thank NOAA/NGDC for making the GOES proton data available and the Oulu Cosmic Ray Station for the GLE list used in this paper. This work benefitted greatly from the open data policy of NASA in using SDO, SAMPEX, SOHO, STEREO, and Wind data. The SWAP data were obtained via the Helioviewer web site. STEREO is a mission in NASA's Solar Terrestrial Probes program. SOHO is a project of international collaboration between ESA and NASA**.** The authors benefitted from a discussion with E. W. Cliver. We thank the anonymous referee for the comments that helped improve the paper. The work of N.G., S.Y., and S.A. was supported by NASA/LWS and H-GI programs. P. M. was partially supported by NSF grant AGS-1358274 and NASA grant NNX15AB77G. H.X. was partially supported by NASA grant NNX15AB70G. N.T. was partially supported by NSF grant AGS-1622377.

**References**

Araki, T. 2014, EP&S, 66,164
Baker, D. N. et al. 1993, IEEE Trans. Geosci. Remote Sens. **31**, 531
Baker, D. N. et al. 2013, Spa Wea 11, 585
Band, D. Matteson, J., Ford, L. et al. 1993, ApJ 413, 281
Bougeret, J.-L., Kaiser, M. L., Kellogg, P. J., et al. 1995, SSRv, 71, 231
Bougeret, J.-L., Goetz, K., Kaiser, M. L. et al. 2008, SSRv 136, 487
Brueckner, G. E., Howard, R. A., Koomen, M. J., et al. 1995, SoPh, 162, 357
Cash, M. D., Biesecker, D. A., Pizzo, V. 2015, SpaWea 13, 611

Cliver, E. W., Feynman, J., & Garrett, H. B. 1990a, Flare-associated solar wind disturbances with short (≤20 hr) transit times to Earth, in SolarTerrestrial Predictions (NOAA: Boulder, CO), vol. 1, edited by R. J. Thompson et al., p. 348
Cliver, E. W., Feynman, J., & Garrett, H. B. 1990b, JGR 95, 17103
Cliver, E. W., Kahler, S. W. & Vestrand, W. T. 1993, 23rd ICRC 3, 91
Cliver, E.W., Nitta, N.V., Thompson, B.J., Zhang, J. 2004, SoPh 225, 105
Cliver, E.W. & Dietrich, W. F. 2013, J. Space Weather Space Clim. 3, A31
Ellison, D. C. & Ramaty, R. 1985, ApJ 298, 400
Gopalswamy, N. 2006 in Solar Eruptions and Energetic Particles (Washington DC: American Geophysical Union Press), Geophysical Monograph Series - Volume 165 p. 207
Gopalswamy, N. 2011, in Proc. 7th Int. Workshop on Planetary, Solar and Heliospheric Radio Emissions (PRE VII), Coronal Mass Ejections and Solar Radio Emissions (Graz: Austrian Academy of Sciences Press), 325
Gopalswamy, N., Lara, A., Kaiser, M.L., Bougeret, J.-L. 2001a, JGR 106, 25261
Gopalswamy, N., Lara, A., Yashiro, S., Kaiser, M.L., Howard, R. A. 2001b, JGR 106, 29207
Gopalswamy, N., Yashiro, S., Liu, Y. et al. 2005a, JGR 110, A09S15
Gopalswamy, N., Aguilar-Rodriguez, E., Yashiro, S., et al. 2005b, JGR, 110, A12S07
Gopalswamy, N., Xie, H., Yashiro, S., Usoskin, I. 2005c, ICRC 1, 169
Gopalswamy, N., Yashiro, S., Michalek, G., et al. 2009a, EM&P 104, 295
Gopalswamy, N., Dal Lago, A., Yashiro, S. & Akiyama, S. 2009b, Cent. Eur. Astrophys. Bull. 33(1), 115
Gopalswamy, N., Thompson, W.T., Davila, D.M. et al. 2009c, SoPh 259, 227
Gopalswamy, N., Xie, H., Yashiro, S., & Usoskin, I. 2010, Indian J. Radio & Space Phys. 39, 240
Gopalswamy, N., Xie, H., Yashiro, S., et al. 2012a, SSRv, 171, 23
Gopalswamy, N., Nitta, N., Akiyama, S., Mäkelä, P., Yashiro, S. 2012b, ApJ 744, 72
Gopalswamy, N., Xie, H., Akiyama, S. et al. 2013a, ApJ 765, L30
Gopalswamy, N., Xie, H., Mäkelä, P., et al. 2013b, AdSpR, 51, 1981
Gopalswamy, N., Xie, H., Akiyama, S., Mäkelä, P., & Yashiro, S. 2014a, EP&S, 66, 104
Gopalswamy N., Akiyama, S., Yashiro, S. et al. 2014b GRL 41, 2673
Gopalswamy, N., Tsurutani, B. T. & Yan, Y. 2015a, Prog. EP&S 2, 13
Gopalswamy, N., Mäkelä, P., Akiyama, S., Yashiro, S., Xie, H., Thakur, N., Kahler, S. W. 2015b ApJ 806, 8
Gopalswamy, N., Yashiro, S., Mäkelä, P., Xie, H., Akiyama, S., & Thakur, N. 2015c, J. Phys. Conf. Ser. 642, 012012, doi:10.1088/1742-6596/642/1/012012
Gopalswamy, N., Yashiro, S., Xie, H., Akiyama, S., & Mäkelä, P. 2015d, JGR 120, 9221
Gopalswamy, N., Xie, H., Akiyama, S., Mäkelä, P., Yashiro, S., & Michalek, G. 2015e, ApJ 804, L23
Howard, R. A., Moses, J. D., Vourlidas, A., et al. 2008, SSRv, 136, 67

Joyce, C. J., Schwadron, N. A., Townsend, L. W. et al. 2015, SpaWea 13, 560
Kahler, S. W. 2001, JGR, 106, 20947
Kaiser, M. L., Kucera, T. A., Davila, J. M. et al. 2008, SSRv 136, 5
Kocharov, L. G., Lee, J. W., Wang, H. et al. 1995, SoPh 158, 95
Lario, D., R. B. Decker, and T. P. Armstrong, Major solar proton events observed by IMP-8 (from November 1973 to May 2001), Proc. 27th ICRC, 3254, 2001.
Lemen J. R., Title A. M., Akin, D. J. et al. 2012, SoPh 275, 17
Liou, K. Wu, C.-C., Dryer, M. et al. 2014, JASTP 121, 32
Liu, Y., Luhmann, J. G., Müller-Mellin, R. et al. 2008, ApJ 689, 563
Liu, Y. D., Luhmann, J. G., Kajdič, P. et al. 2014, Nature Communications 5, 3481
Mäkelä, P. Gopalswamy, N., Akiyama, S., Xie, H., Yashiro, S. 2011, ApJ 116, A08101
Mäkelä, P., Gopalswamy, N., Akiyama, S., Xie, H., & Yashiro, S. 2015, ApJ 806, 13
Mann, G., Klassen, A., Aurass, H., & Classen, H.-T. 2003, A&A 400,329
McGuire, R. E. & von Rosenvinge, T. T. 1984, AdSpR 4, 117
Mewaldt, R. A., Cohen, C. M. S., Labrador, A. W. et al. 2005, JGR 110, 9S18M
Mewaldt, R. A., Looper, M. D., Cohen, C. M. S. et al. 2012, SSRv 171, 97
Mewaldt, R.A., C.T. Russell, C.M.S. Cohen, A.B. Galvin, R. Gomez-Herrero, A. Klassen, R.A. Leske, J. Luhmann, G.M. Mason, and T.T. von Rosenvinge, A 360 view of solar energetic particle events, including one extreme event, in: Proc. 33rd Int. Cosmic Ray Conf., 2013, p. 1186
Miroschnichenko, L. I., de Koning, C. A., & Perez-Enriquez, R. 2000, SSRv 91, 615
Ngwira, C., Pulkkinen, A., Mays, L. et al. 2013, Spa Wea 11, 671
Reames, D. V. 1999, SSRv, 90, 413
Reames, D. V. 2009a, ApJ 693, 812
Reames, D. V. 2009b, ApJ 706, 844
Reames, D. V. 2013, SSRv, 175, 53
Reeves, G. D., Cayton, T. E., Gary, S. P., Belian, R. D. 1992, JGR 97, 6219
Riley, P., Caplan, R. M., Giacalone, J., Lario, D., & Liu, Y. 2016, ApJ 819, 57
Russell, C. T., Mewaldt, R. A., Luhmann, J. G. et al. 2013, ApJ 770, 42
Sandberg, I., Jiggens, P., Heynderickx, D., & Daglis, I. A. 2014, GRL. 41, 4435
Sarris, E. T., Anagnostopoulos, G. C., Trochoutsos, P. C. 1984, SoPh 93, 195
Seaton, D. B., Berghmans, D., Nicula, B. et al. 2013, SoPh 286, 43
Shea, M. A. & Smart, D. F. 1993, 23rd ICRC 3, 739
Shea, M. A. & Smart, D. F. 2012, SSRv 171, 161
Temmer, M & Nitta N. V. 2015, SoPh 290, 919
Thakur, N., Gopalswamy, N., Xie, H., et al. 2014, ApJ 790, L13
Thakur, N., Gopalswamy, N., Mäkelä, P. et al. 2016, SoPh 291, 513
Thernisien, A. 2011, ApJS, 194, 33
Thompson, R., Kennewell, J., & Prestage, N. 1996, SoPh 166, 371
Torsti, J., Valtonen, E., Lumme, M., et al. 1995, SoPh, 162, 505

Usoskin, I. G., Mursula, K., Kangas, J. 2001, On-Line Database of Cosmic Ray Intensities, Proc. 27th International Cosmic Ray Conference. 07-15 August, 2001. Hamburg, Germany. Under the auspices of the International Union of Pure and Applied Physics (IUPAP), p.3842
von Rosenvinge, T. T. et al. 2008, SSRv136, 391
Vršnak, B., Žic, T., Vrbanec, D. et al. 2013, SoPh 285, 295
Yashiro, S., Gopalswamy, N., Michalek, G., et al. 2004, JGR 109, A07105
Zhang, J., Kundu, M. R., White, S. M., Dere, K. P., Newmark, J. S. 2001, ApJ 561, 396
Zhang, J., & Dere, K. 2006, ApJ, 649, 1100
Zhu, B., Liu, Y. D., Luhmann, J. G. et al. 2016, ApJ 827, 146
Zurbuchen, T. H., Raines, J. M., Slavin, J. A., et al. 2011, Sci, 333, 1862

Table 2. List of SEP events, the fluence spectral indices, and the CME information

| **SEP DATE** | **TIME** | **SEP TYPE** | $\gamma$ | $\gamma_{err}$ | **E Range (MeV)** | **CME TIME** | $V_{in}$ | $V_{sky}$ | $V_{Sp}$ | **Location** | **Flare Size** | **Flare Start** | **Flare Peak** | $a_{in}$ |
|---|---|---|---|---|---|---|---|---|---|---|---|---|---|---|
| 1997/11/04 | 06:40 | SEP | 2.82 | 0.09 | 20.0-80.6 | 06:10 | 1320 | 785 | 993 | S14W33 | X2.1 | 05:52 | 05:58 | 2.76 |
| 1997/11/06 | 12:30 | GLE | 2.07 | 0.08 | 20.0-80.6 | 12:10 | 1717 | 1556 | 1604 | S18W63 | X9.4 | 11:49 | 11:55 | 4.46 |
| 1998/04/20 | 11:15 | SEP | 3.23 | 0.15 | 20.0-80.6 | 10:07 | 1543 | 1863 | 1863 | S43W90 | M1.4 | 09:38 | 10:21 | 0.72 |
| 1998/05/02 | 14:00 | GLE | 2.01 | 0.09 | 20.0-80.6 | 14:06 | 1585 | 938 | 1168 | S15W15 | X1.1 | 13:31 | 13:42 | 1.77 |
| 1998/05/06 | 08:25 | GLE | 2.67 | 0.12 | 20.0-80.6 | 08:29 | 1052 | 1099 | 1210 | S11W65 | X2.7 | 07:58 | 08:09 | 1.83 |
| 1998/05/09 | 06:20 | SEP | 3.31 | 0.23 | 20.0-59.1 | 03:35 | 2880 | 2331 | 2331 | S11W90 | M7.7 | 03:16 | 03:40 | 1.62 |
| 1998/08/24 | 23:10 | GLE | 3.79 | 0.12 | 20.0-80.6 | ... | ... | ... | ... | ... | ... | ... | ... | ... |
| 1998/09/30 | 14:20 | SEP | 4.53 | 0.16 | 20.0-80.6 | ... | ... | ... | ... | ... | ... | ... | ... | ... |
| 1998/11/14 | 06:30 | SEP | 2.64 | 0.12 | 20.0-80.6 | ... | ... | ... | ... | ... | ... | ... | ... | ... |
| 1999/06/04 | 08:30 | SEP | 4.64 | 0.25 | 20.0-59.1 | 07:26 | 2335 | 2230 | 2373 | N17W69 | M3.9 | 06:52 | 07:03 | 3.60 |
| 2000/04/04 | 17:00 | FE | 5.56 | 0.16 | 20.0-59.1 | 16:32 | 1107 | 1188 | 1217 | N16W66 | C9.7 | 15:12 | 15:41 | 0.70 |
| 2000/06/10 | 17:40 | SEP | 2.77 | 0.09 | 20.0-80.6 | 17:08 | 1297 | 1108 | 1228 | N22W38 | M5.2 | 16:40 | 17:02 | 0.93 |
| 2000/07/14 | 10:35 | GLE | 2.98 | 0.19 | 20.0-59.1 | 10:54 | 1746 | 1674 | 2061 | N22W07 | X5.7 | 10:03 | 10:24 | 1.64 |
| 2000/07/22 | 12:25 | SEP | 3.30 | 0.12 | 20.0-67.8 | 11:54 | 1129 | 1230 | 1476 | N14W56 | M3.7 | 11:17 | 11:34 | 1.45 |
| 2000/09/12 | 13:50 | FE | 3.81 | 0.08 | 20.0-80.6 | 11:54 | 937 | 1550 | 1946 | S19W06 | M1.0 | 11:23 | 12:13 | 0.65 |
| 2000/10/16 | 08:10 | SEP | 3.23 | 0.09 | 20.0-80.6 | 07:27 | 1195 | 1336 | 1336 | N03W90 | M2.5 | 06:40 | 07:28 | 0.46 |
| 2000/10/25 | 13:35 | FE | 4.51 | 0.20 | 20.0-59.1 | 08:26 | 307 | 770 | 813 | N10W66 | C4.0 | 08:45 | 11:25 | 0.08 |
| 2000/11/08 | 23:35 | SEP | 2.79 | 0.09 | 20.0-80.6 | 23:06 | 1137 | 1738 | 1783 | N10W77 | M7.4 | 23:04 | 23:28 | 1.24 |
| 2001/01/28 | 17:15 | SEP | 3.55 | 0.13 | 20.0-80.6 | 15:54 | 463 | 916 | 1068 | S04W59 | M1.5 | 15:40 | 16:00 | 0.89 |
| 2001/04/02 | 23:10 | SEP | 3.37 | 0.14 | 20.0-80.6 | 22:06 | 2652 | 2505 | 2609 | N19W72 | X20. | 21:32 | 21:51 | 2.29 |
| 2001/04/12 | 11:40 | SEP | 2.66 | 0.14 | 20.0-80.6 | 10:31 | 1462 | 1184 | 1294 | S19W43 | X2.0 | 10:13 | 10:28 | 1.44 |
| 2001/04/15 | 14:00 | GLE | 2.33 | 0.14 | 20.0-80.6 | 14:06 | 1310 | 1199 | 1203 | S20W85 | X14. | 13:42 | 13:50 | 2.51 |
| 2001/04/18 | 03:00 | GLE | 2.59 | 0.15 | 20.0-80.6 | 02:30 | 2664 | 2465 | 2465 | S17W120 | ... | ... | ... | ... |
| 2001/08/09 | 19:30 | FE | 5.37 | 0.34 | 20.0-41.6 | 10:30 | 328 | 479 | 683 | N11W14 | ... | ... | ... | ... |
| 2001/09/15 | 12:05 | SEP | 3.90 | 0.20 | 20.0-59.1 | 11:54 | 630 | 478 | 586 | S21W49 | M1.5 | 11:04 | 11:28 | 0.41 |
| 2001/10/01 | 12:15 | SEP | 5.28 | 0.22 | 20.0-80.6 | 05:30 | 642 | 1405 | 1409 | S20W84 | M9.1 | 04:41 | 05:15 | 0.69 |
| 2001/10/19 | 18:00 | SEP | 3.06 | 0.09 | 20.0-80.6 | 16:50 | 1355 | 901 | 1051 | N15W29 | X1.6 | 16:21 | 16:30 | 1.95 |
| 2001/11/04 | 16:40 | GLE | 2.98 | 0.15 | 20.0-80.6 | 16:35 | 2071 | 1810 | 2286 | N06W18 | X1.0 | 16:03 | 16:20 | 2.24 |
| 2001/11/23 | 01:05 | SEP | 4.43 | 0.1 | 20.0-80.6 | 23:30[a] | 1923 | 1437 | 1577 | S17W36 | M9.9 | 22:28 | 23:30 | 0.42 |
| 2001/12/26 | 05:50 | GLE | 2.67 | 0.08 | 20.0-80.6 | 05:30 | 1624 | 1406 | 1723 | N08W54 | M7.1 | 04:32 | 05:40 | 0.42 |
| 2002/01/15 | 05:35 | SEP | 4.14 | 0.14 | 20.0-80.6 | 05:35[a] | 488 | 1492 | 1492 | S28W90 | M4.4 | 05:29 | 06:27 | 0.43 |
| 2002/02/20 | 06:45 | SEP | 4.02 | 0.23 | 20.0-59.1 | 06:30 | 1237 | 952 | 965 | N12W72 | M5.1 | 06:06 | 06:12 | 2.68 |
| 2002/03/18 | 06:45 | SEP | 4.93 | 0.26 | 20.0-59.1 | 02:54 | 948 | 989 | 1223 | S09W46 | M1.0 | 02:16 | 02:31 | 1.36 |
| 2002/03/22 | 13:50 | SEP | 5.87 | 0.47 | 20.0-32.4 | 11:06 | 1643 | 1750 | 1750 | S09W90 | M1.6 | 10:34 | 11:14 | 0.73 |
| 2002/04/17 | 11:35 | SEP | 4.78 | 0.25 | 20.0-59.1 | 08:26 | 1181 | 1218 | 1417 | S14W34 | M2.6 | 07:46 | 08:24 | 0.62 |
| 2002/04/21 | 01:45 | SEP | 2.77 | 0.12 | 20.0-80.6 | 01:27 | 2123 | 2409 | 2422 | S14W84 | X1.5 | 00:43 | 01:51 | 0.59 |
| 2002/05/22 | 07:10 | FE | 4.55 | 0.16 | 20.0-67.8 | 03:50 | 1586 | 1494 | 1718 | S30W34 | C5.0 | 03:18 | 03:54 | 0.80 |
| 2002/07/07 | 12:45 | SEP | 3.76 | 0.24 | 20.0-59.1 | 11:30 | 1337 | 1423 | 1423 | S19W90 | M1.0 | 11:15 | 11:43 | 0.85 |
| 2002/08/14 | 02:50 | SEP | 4.91 | 0.20 | 20.0-41.6 | 02:30 | 1353 | 1309 | 1617 | N09W54 | M2.3 | 01:47 | 02:12 | 1.08 |
| 2002/08/22 | 02:40 | SEP | 2.45 | 0.10 | 20.0-80.6 | 02:06 | 1442 | 1005 | 1034 | S07W62 | M5.4 | 01:47 | 01:57 | 1.72 |
| 2002/08/24 | 01:30 | GLE | 2.87 | 0.10 | 20.0-80.6 | 01:27 | 2073 | 1878 | 1920 | S02W81 | X3.1 | 00:49 | 01:12 | 1.39 |
| 2002/11/09 | 15:25 | SEP | 5.43 | 0.32 | 20.0-59.1 | 13:31 | 1894 | 1838 | 2159 | S12W29 | M4.6 | 13:08 | 13:23 | 2.40 |
| 2003/05/28 | 04:15 | SEP | 4.74 | 0.24 | 20.0-59.1 | 00:50 | 1092 | 1366 | 1701 | S07W20 | X3.6 | 00:17 | 00:27 | 2.83 |
| 2003/05/31 | 03:00 | SEP | 2.45 | 0.13 | 20.0-80.6 | 02:30 | 1838 | 1835 | 1888 | S07W65 | M9.3 | 02:13 | 02:24 | 2.86 |
| 2003/10/26 | 18:00 | SEP | 4.38 | 0.16 | 20.0-80.6 | 17:54 | 1044 | 1537 | 2491 | N02W38 | X1.2 | 17:21 | 18:19 | 0.72 |
| 2003/10/28 | 11:45 | GLE | 2.66 | 0.71 | 20.0-32.4 | 11:30 | 2832 | 2459 | 3128 | S16E08 | X17. | 11:00 | 11:10 | 5.21 |
| 2003/10/29 | 21:55 | GLE | 2.83 | 0.11 | 20.0-80.6 | 20:54 | 2264 | 2020 | 2628 | S15W02 | X10. | 20:37 | 20:49 | 3.65 |
| 2003/11/02 | 17:35 | GLE | 3.50 | 0.12 | 20.0-80.6 | 17:30 | 2635 | 2598 | 2733 | S14W56 | X8.3 | 17:03 | 17:25 | 2.07 |
| 2003/11/04 | 22:15 | SEP | 3.65 | 0.12 | 20.0-80.6 | 19:54 | 2344 | 2657 | 2662 | S19W83 | X28. | 19:38 | 19:53 | 2.96 |
| 2004/04/11 | 06:10 | FE | 4.50 | 0.13 | 20.0-59.1 | 04:30 | 1869 | 1645 | 2230 | S14W47 | C9.6 | 03:54 | 04:19 | 1.49 |
| 2004/07/25 | 16:55 | SEP | 5.98 | 0.32 | 20.0-59.1 | 14:54 | 957 | 1333 | 1544 | N08W33 | M1.1 | 14:19 | 15:14 | 0.47 |
| 2004/09/19 | 18:10 | SEP | 3.92 | 0.21 | 20.0-67.8 | ... | ... | ... | ... | ... | ... | ... | ... | ... |
| 2005/01/17 | 12:40 | GLE | 3.02 | 0.14 | 20.0-80.6 | 09:54 | 2938 | 2547 | 3029 | N15W25 | X3.8 | 09:42 | 09:52 | 5.05 |
| 2005/01/20 | 06:55 | GLE | 2.13 | 0.08 | 20.0-80.6 | 06:54 | 853 | 882 | 951 | N14W61 | X7.1 | 06:36 | 07:01 | 0.63 |
| 2005/06/16 | 20:45 | SEP | 2.17 | 0.10 | 20.0-80.6 | ... | ... | ... | ... | ... | ... | ... | ... | ... |
| 2005/07/14 | 13:05 | SEP | 4.87 | 0.22 | 20.0-80.6 | 10:54 | 1971 | 2115 | 2115 | N11W90 | X1.2 | 10:16 | 10:55 | 0.90 |
| 2005/08/22 | 19:20 | SEP | 5.28 | 0.15 | 20.0-80.6 | 17:30 | 1908 | 2378 | 2445 | S13W65 | M5.6 | 16:46 | 17:27 | 0.99 |
| 2006/12/13 | 02:55 | GLE | 2.07 | 0.08 | 20.0-80.6 | 02:54 | 1868 | 1774 | 2184 | S06W23 | X3.4 | 02:14 | 02:40 | 1.40 |
| 2006/12/14 | 22:55 | SEP | 2.60 | 0.08 | 20.0-80.6 | 22:30 | 1154 | 1042 | 1139 | S06W46 | X1.5 | 21:58 | 22:15 | 1.12 |
| 2010/08/14 | 11:05 | SEP | 3.90 | 0.39 | 12.5-46.1 | 10:12 | 1216 | 1205 | 1658 | N17W52 | C4.4 | 09:38 | 10:05 | 1.02 |
| 2011/03/07 | 21:45 | SEP | 4.70 | 0.23 | 12.5-104. | 20:00 | 1843 | 2125 | 2660 | N31W53 | M3.7 | 19:43 | 20:12 | 1.53 |
| 2011/03/21 | 04:10 | SEP | 3.43 | 0.33 | 12.5-104. | 02:24 | 1216 | 1341 | 1430 | ... | ... | ... | ... | ... |
| 2011/06/07 | 07:20 | SEP | 2.22 | 0.22 | 12.5-104. | 06:49 | 1654 | 1255 | 1680 | S21W54 | M2.5 | 06:16 | 06:41 | 1.12 |
| 2011/08/04 | 04:30 | SEP | 3.62 | 0.16 | 12.5-104. | 04:12 | 1546 | 1315 | 2450 | N19W36 | M9.3 | 03:49 | 03:57 | 5.10 |
| 2011/08/09 | 08:20 | SEP | 2.50 | 0.28 | 12.5-104. | 08:12 | 2200 | 1610 | 2496 | N17W69 | X6.9 | 07:59 | 08:05 | 6.93 |
| 2011/11/26 | 08:15 | FE | 5.46 | 0.20 | 12.5-104. | 07:12 | 696 | 933 | 1187 | N17W49 | C1.2 | 06:09 | 07:10 | 0.32 |
| 2012/01/27 | 18:55 | SEP | 3.32 | 0.35 | 12.5-104. | 18:27 | 1817 | 2508 | 2625 | N27W71 | X1.7 | 18:03 | 18:37 | 1.29 |
| 2012/03/13 | 18:05 | SEP | 3.73 | 0.19 | 12.5-104. | 17:36 | 1927 | 1884 | 2333 | N17W66 | M7.9 | 17:12 | 17:41 | 1.34 |
| 2012/05/17 | 01:55 | GLE | 2.48 | 0.12 | 12.5-104. | 01:48 | 1756 | 1582 | 1997 | N11W76 | M5.1 | 01:25 | 01:47 | 1.51 |
| 2012/07/07 | 00:05 | SEP | 3.50 | 0.21 | 12.5-104. | 23:24[a] | 3276 | 1828 | 2464 | S13W59 | X1.1 | 23:01 | 23:08 | 5.87 |
| 2012/07/08 | 18:10 | SEP | 2.01 | 0.60 | 12.5-104. | 16:54 | 2078 | 1495 | 2905 | S17W74 | M6.9 | 16:23 | 16:32 | 5.38 |
| 2012/07/12 | 17:25 | SEP | 6.12 | 0.24 | 12.5-104. | 16:48 | 617 | 885 | 1415 | S15W01 | X1.4 | 16:00 | 16:49 | 0.48 |
| 2012/07/17 | 15:30 | SEP | 5.18 | 0.42 | 12.5-104. | 13:48 | 261 | 958 | 1881 | S15W65 | C9.9 | 13:18 | 15:59 | 0.19 |
| 2012/07/19 | 06:40 | SEP | 3.48 | 0.54 | 12.5-104. | 05:24 | 1003 | 1631 | 2048 | S13W88 | M7.7 | 05:07 | 05:58 | 0.67 |
| 2012/09/28 | 01:20 | SEP | 4.00 | 0.33 | 12.5-104. | 00:12 | 1256 | 947 | 1479 | N06W34 | C3.7 | 23:36 | 23:57 | 1.17 |
| 2013/05/22 | 14:20 | SEP | 3.78 | 0.41 | 12.5-104. | 13:25 | 1529 | 1466 | 1881 | N15W70 | M5.0 | 12:58 | 13:32 | 0.92 |
| 2013/09/30 | 00:25 | SEP | 5.14 | 0.24 | 12.5-104. | 22:12[a] | 946 | 1179 | 1864 | N23W25 | C1.3 | 21:43 | 23:39 | 0.27 |
| 2013/12/28 | 19:00 | SEP | 3.91 | 0.37 | 12.5-104. | 17:36 | 949 | 1118 | 1918 | ... | ... | ... | ... | ... |
| 2014/01/06 | 08:15 | GLE | 2.54 | 0.11 | 12.5-104. | 08:00 | 1843 | 1402 | 2287 | ... | ... | ... | ... | ... |
| 2014/01/07 | 19:55 | SEP | 4.27 | 0.55 | 12.5-104. | 18:24 | 2348 | 1830 | 3121 | S12W11 | X1.2 | 18:04 | 18:32 | 1.86 |
| 2014/02/20 | 08:15 | SEP | 2.87 | 0.49 | 12.5-104. | 08:00 | 1124 | 948 | 1281 | S15W73 | M3.0 | 07:26 | 07:56 | 0.71 |
| 2014/04/18 | 13:40 | SEP | 4.15 | 0.05 | 12.5-104. | 13:25 | 1438 | 1203 | 1711 | S20W34 | M7.3 | 12:31 | 13:03 | 0.89 |
| 2014/11/01 | 13:55 | FE | 5.53 | 0.67 | 12.5-46.1 | 06:00 | 897 | 740 | 1000 | N27W79 | EP | ... | ... | ... |
| 2015/06/18 | 04:35 | SEP | 4.94 | 0.98 | 12.5-46.1 | 01:25 | 2176 | 1714 | 1733 | S16W81 | M1.2 | 00:33 | 01:27 | 0.53 |
| 2015/10/29 | 03:05 | SEP | 2.28 | 0.18 | 12.5-104. | 02:36 | 734 | 530 | 530 | ... | ... | ... | ... | ... |
| 2016/01/02 | 00:15 | SEP | 3.99 | 0.22 | 12.5-104. | 23:24[a] | 1130 | 1760 | 1774 | S25W82 | M2.3 | 23:10[a] | 00:11 | 0.48 |

**Notes**. No entries (…) are due to LASCO data gap or behind the west limb events; [a]Time corresponds to the previous day

$V_{in}$ – CME speed in km $s^{-1}$ obtained from the first two height-time measurements in LASCO/C2 FOV

$V_{Sky}$ – Average CME speed in km $s^{-1}$ in the LASCO/C2 FOV

$V_{Sp}$ – CME space speed in km $s^{-1}$ in LASCO/C2 FOV deprojected using the cone model or geometric deprojection if the eruption occurred within 30º from the limb; for the 2001 August 09 event, we used the empirical relation $V_{Sp}$= 1.1*$V_{Sky}$+156 (Gopalswamy et al. 2015e).